# Evolution of Research Method Usage Across the Academic Careers of Library and Information Science Scholars

Jiayi Hao and Chengzhi Zhang[*]

Nanjing University of Science and Technology, Department of Information Management, Nanjing, 210094 (China)

**Abstract**: Research methods constitute an indispensable tool for scholars engaged in scientific inquiry. Investigating how scholars use research methods throughout their careers can reveal distinct patterns in method adoption, providing valuable insights for novice researchers in selecting appropriate methods. This study employs a comprehensive dataset comprising full-text journal articles and bibliographic records from the Library and Information Science (LIS) domain. Utilizing an automated classification model based on full-text cognitive analysis, the research methods employed by LIS scholars are systematically identified. Topic modeling was then conducted using Top2Vec. Subsequently, author name disambiguation is performed, and academic age is calculated for each scholar. This study focuses on 435 senior scholars with an academic age of more than 14 years and a consistent publication record at five-year intervals, covering a total of 6,116 articles. The corpus covers 16 research method categories and 20 research topics. The findings indicate that bibliometric methods are the most frequently used across career stages, accounting for 19.61% among early-career scholars and 31.81% among senior scholars. Over the course of a scholarly career, the diversity of research methods initially increases and then declines. Furthermore, scholars exhibit a propensity for combining multiple research methods, including both conventional and unconventional pairings. Notably, the research methods most commonly used by researchers change with age and seniority.

## 1 Introduction

Researchers' scholarly endeavors drive scientific progress. Studying scholars themselves offers valuable insights into the mechanisms shaping modern science. Age, a key scholar attribute, significantly influences academic performance. As scholars age, their cognitive abilities and perspectives expand (Wang et al., 2017), shaping their research interests, methodol choices, and output.

Given the unique and complex nature of academic research, prior studies have adopted the lens of academic age to more precisely delineate and comprehend the developmental evolution and stage-specific characteristics of scholars within their respective fields. Academic age is typically calculated based on the timing of a scholar's first publication (Costas et al., 2015). This metric correlates with research productivity (Abramo et al., 2016; Győrffy et al., 2020), academic influence (Sugimoto et al., 2016), and collaborative networks (Bu et al., 2018; Kumar & Ratnavelu, 2016; Wang et al., 2017). Understanding how scholars select and shift their research focus over time is crucial for scientist training, funding allocation, knowledge organization, and reward systems (Jia et al., 2017). Academic age also helps distinguish career stages. Empirical studies show that as academic age increases, scholars accumulate more resources, explore more diverse topics, and tend to be more productive (Abramo et al., 2016; Simoes & Crespo, 2020; Zeng et al., 2019). However, disparities exist across age groups. While senior scholars benefit from experience, funding, and collaboration, their knowledge base often stabilizes later in their careers. This can lead to reliance on outdated concepts (Liang et al., 2020; Milojević, 2012; Packalen & Bhattacharya, 2019), reduced receptivity to new ideas (Azoulay et al., 2019), and engagement in less prominent research areas (Cui et al., 2022). Consequently, scholars at different stages of their academic careers exhibit distinct cognitive behaviors and research patterns.

Research methods serve as cognitive frameworks guiding scientific inquiry and are essential in any discipline. As a cornerstone of research, their significance and need for innovation have become increasingly evident. Studies have revealed notable age-related differences in the research methods employed by scholars at various stages of their academic careers. Senior scholars exhibit a predilection for qualitative research, while their junior counterparts tend to favor quantitative methodologies (Lou et al., 2021). This discrepancy reflects the phased characteristics of scholars' academic cognition and research experience. For junior scholars entering the field, they often face confusion in selecting research methods: they are uncertain about which methods to prioritize at their

---

current stage and lack access to the method expertise of senior scholars. As a result, they may blindly follow quantitative research trends while neglecting the alignment between methods and research questions. Therefore, it is necessary to explore the evolution of method choices throughout scholars' academic careers, providing young scholars with a clear reference framework for method selection. This would help them plan their methodological learning and application pathways more effectively, avoiding inefficiencies caused by inappropriate method choices. Moreover, prior studies have primarily examined the effects of academic age from perspectives such as team collaboration and academic output, or have focused on the classification, identification, and application of research methods in academic papers. Few have integrated scholars' academic age with their method practices to conduct a comprehensive examination of how research methods evolve over the course of an academic career. This study distinguishes itself from prior research in several key aspects. It not only clarifies the research methods used in the LIS field based on large-scale literature data, but also uses deep learning models to identify method usage within this extensive corpus. Furthermore, it integrates scholars' academic career evolution to examine how their method practices evolve over time. Based on this approach, the study can reveal characteristic patterns in method adoption across different stages of a scholarly career. These insights offer practical value by helping young scholars quickly grasp the method landscape of the field. Additionally, the findings can also help universities or research institutions design targeted training courses on research methods, providing young scholars with guidance in selecting research methods and helping to enhance their research capabilities.

This study uses journal literature to investigate the evolution of research method usage among LIS scholars, addressing two research questions:

*RQ1:* What differences exist in the research methods used by LIS scholars at different stages of their academic age?

*RQ2:* What patterns characterize the evolution of research method usage throughout LIS scholars' academic careers?

## 2    Literature review

This paper explores the evolution of research method selection in scholars' academic careers within a specific field. Due to limited research on academic evolution, we focus on two dimensions: academic age and research method use.

### 2.1    Academic age of scholars in specific fields

Research on academic age involves its definition and various research dimensions.

Academic age is typically calculated using either a scholar's first publication date or doctoral graduation year, with studies varying in scale. However, the scale of these studies varies significantly. Research utilizing the first publication date to determine academic age encompasses a wide range of sample sizes. Smaller-scale studies span diverse fields, such as 137 scholars in information systems (Liao, 2017) and 472 top economists (Simoes & Crespo, 2020). Larger-scale studies include 21,562 scientists across five disciplines and ten core journals (Milojević, 2012), 94,000 scientists from 43 countries (Chan & Torgler, 2020), and even 222,925 authors (Robinson-Garcia et al., 2020) or 1.7 million author records from the Web of Science platform (Aref et al., 2019). In contrast, studies using the year of doctoral graduation to calculate academic age typically involve smaller samples, often numbering in the hundreds (Badar et al., 2014; Chan & Torgler, 2020; Coomes et al., 2013) or thousands (Perianes-Rodriguez & Ruiz-Castillo, 2015; Sugimoto et al., 2016). For instance, van den Besselaar and Sandström (2016) examined 243 researchers applying for early-career grants in the Netherlands, while Perianes-Rodriguez and Ruiz-Castillo (2015) analyzed 2,530 economists working in 81 top global economics departments. Costas et al. (2015) utilized a real-world dataset of professors in Quebec to evaluate the feasibility of these two metrics and concluded that the first publication date is a more suitable indicator of a researcher's academic age. Similarly, Nane et al. (2017) identified the year of first publication as the best linear predictor of a scholar's age. Therefore, this study defines the start of an academic career as the date of first publication.

Academic age is commonly studied alongside scholars' academic or career stages, with investigations spanning multiple domains, as illustrated in Table 1. From an application perspective, academic age has been widely used to study various research behaviors. Most existing studies examine teamwork, career dynamics, and scientific networks by analyzing data from scientometrics and network science to interpret scientists' cognitive processes and behavioral patterns (Krauss, 2024).

**Table 1.** Different research perspectives integrating scholars' academic careers

| Authors | Perspective | Main findings |
| --- | --- | --- |
| Milojević (2012) | Reference citation behaviour | Similar citation behavior with senior and junior researchers citing references at comparable rates and consistent re-citation patterns |
| Aref et al. (2019) | Researcher mobility | Hypermobility analysis categorizing scholars at early mid and late career stages by academic age and identifying destination countries |
| Simoes and Crespo (2020) | Performance assessment | Publication productivity showing longer careers linked to higher output and prolific authorship |
| Robinson-Garcia et al. (2020) | Career stages | Career stage biases revealed through academic age and author contribution statements indicating variations in scientific career stages |
| Ao et al. (2023) | Patterns of scientific creativity | Disruption index trends with both male and female scholars showing a "high peak" creativity pattern and a small subset of females exhibiting an "early peak" |
| Zhang et al. (2024) | Changes in research direction | Research direction shifts with women changing direction less frequently than men and experiencing less negative performance impact |

## 2.2 Overview of research on the use of research methods in specific fields

Research methods, as a defining feature of scholarly activity, represent a core pathway to understanding the mechanisms of scientific practice (Liu et al., 2023). Within the LIS field, scholars have examined the current status and trends of research method usage from various perspectives. As shown in Table 2, existing studies primarily focus on three directions. First, the construction of classification systems for research methods. Järvelin & Vakkari (1990) categorized methods in core LIS journals into 9 research strategies and 10 data collection techniques. Chu (2015) focused on data collection, categorizing LIS methods into 16 types. Second, the application trends of mixed methods. Hayman and Smith (2020) analyzed the use of mixed methods in articles, examining the extent of mixed methods research in LIS over the past decade (2008–2018) and the volume of such studies in health-related contexts. Third, the overall evolutionary patterns of research methods. Lund & Wang (2021) visualized the growth in methodological diversity in LIS from 1980 to 2019. Lou et al. (2021) investigated how researchers in different age groups employ research methods over time. Järvelin and Vakkari (2021) expanded on their earlier work by summarizing the methodological evolution in LIS over the past 50 years, noting that LIS research has become increasingly methodologically diverse, with more varied approaches to analyzing research subjects. Zhang et al. (2023) conducted a longitudinal study on the frequency and diversity of research methods in LIS, revealing a shift from conceptual to empirical research strategies over 31 years.

In summary, the considerable attention scholars have paid to research method usage has contributed to refining method paradigms in the field. However, existing LIS research has not yet deeply linked a scholar's academic career with the research behavior of method selection. This gap is precisely what the present study aims to fill. Therefore, this paper will incorporate academic age as an indicator to investigate how research methods evolve among scholars at different career stages.

**Table 2.** Studies on the use of research methods

| Authors | Perspective | Main findings |
|---|---|---|
| Järvelin and Vakkari (1990) | Classification of research methods | Systematic categorization of research methods into 9 strategies and 10 data collection techniques |
| Chu (2015) | Classification of research methods | LIS research methods classified into 16 categories based on data collection |
| Hayman and Smith (2020) | Use of mixed research methods | Mixed methods in LIS showing small but significant growth over the past decade |
| Lund and Wang (2021) | Changing trends in the use of various research methods. | Increasing method diversity with data analysis and qualitative methods dominating recent publications |
| Lou et al. (2021) | Researchers in different age groups use research methods over time | Rise in quantitative methods driven by younger researchers and senior scholars |
| Järvelin and Vakkari (2021) | Research evolution in the field of LIS | Methodological fragmentation in LIS over 50 years reflecting diversified analytical approaches. |
| Zhang et al. (2023) | Frequency and diversity of application of research methods | Shift in LIS research strategies from conceptual to empirical over 31 years |

# 3 Data and methodology

This section outlines the research framework and key steps for investigating the evolution of research method selection throughout scholars' academic careers. The framework includes data sources, research method classification, and the acquisition of academic career data.

## 3.1 Framework

This study aims to investigate the evolution of method choices throughout the academic careers of scholars. First, full-text journal articles and relevant metadata from the field serve as the data source. Machine learning techniques are employed to identify the research methods used in these publications. Second, topic modeling is applied to article abstracts to extract prevalent research topics. Furthermore, author name disambiguation is performed on the publication set to distinguish individual scholars. Academic age and other scholar-specific metrics are subsequently calculated. Finally, data from selected senior scholars are analyzed to examine longitudinal patterns in their selection of research methods. The research framework is shown in Figure 1.

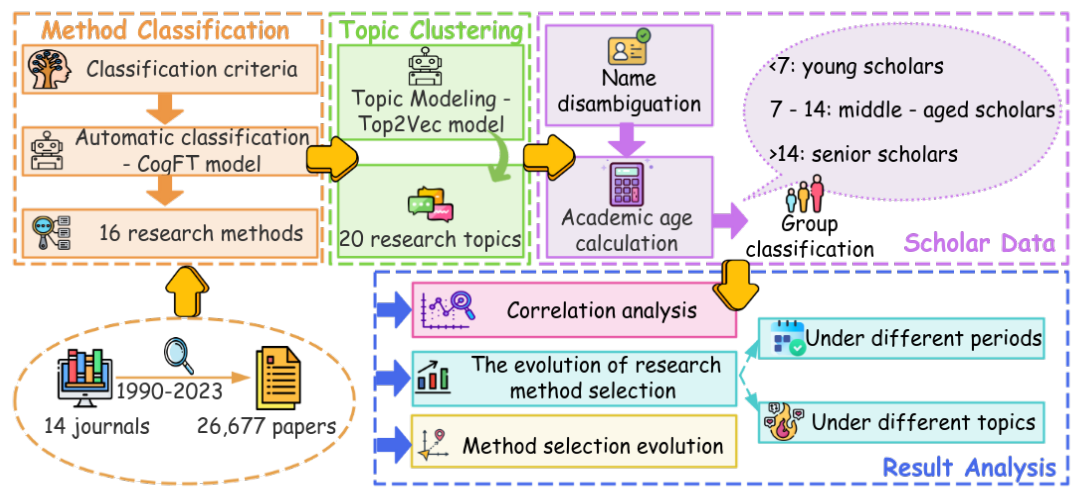

**Figure 1.** Framework of this study

## 3.2 Data sources

To ensure data quality, this study selected representative academic journals from the LIS field. In prior research, Järvelin and Vakkari (1993) conducted extensive studies on research methods and identified 31 representative academic journals in LIS based on the research topics covered in their articles. Building on this foundation, this study integrates the list of representative journals identified by Järvelin and colleagues with the 2023 Journal Citation Reports (JCR) LIS category, which includes core journals across quartiles Q1 to Q4. This process resulted in the selection of 14 high-quality, representative LIS journals. Data collection included both metadata and full-text content from 1990 to 2023. Full-text data were obtained from the official websites of each journal and converted into Word document format using conversion tools. These documents were then processed and parsed using Python to generate standardized full-text data. For cases where metadata were incomplete, bibliographic data for all articles published in the 14 journals over the 34-year period were downloaded from the Web of Science (WoS; https://www.webofscience.com), and missing metadata were supplemented using DOI matching. In total, this study compiled full-text and metadata for 26,677 academic articles published in LIS journals between 1990 and 2023. The number of articles per journal is detailed in Table 3.

Among the 14 journals, the three journals with the highest number of data entries are *Scientometrics*, *Journal of the Association for Information Science and Technology*, and *Information Processing & Management*. These journals collectively account for 12,917 articles, representing nearly 50% of the total dataset.

## 3.3 Classification of research methods for academic papers in the LIS field

Based on the constructed full text corpus of academic papers in the field of LIS, this study classifies and identifies the research methods employed in these articles. The process involves two main steps. Firstly, a suitable classification system of research methods is selected. Secondly, based on the classification system, a technique of automatic classification of research methods is used to identify the research methods of academic papers in the corpus and obtain the results of classification of research methods.

**Table 3.** Number of academic articles in high quality representative journals in the field of LIS

| Journal name | Abbreviation | #Articles |
|---|---|---|
| *Aslib Journal of Information Management* | *AJIM* | 1356 |
| *College & Research Libraries* | *CRL* | 1330 |
| *Information Processing & Management* | *IPM* | 3063 |
| *Information Technology and Libraries* | *ITL* | 546 |
| *International Journal of Information Management* | *IJIM* | 1891 |
| *Journal of Documentation* | *JOD* | 1450 |
| *Journal of Information Science* | *JIS* | 1510 |
| *Journal of Librarianship and Information Science* | *JLIS* | 887 |
| *Journal of the Association for Information Science and Technology* | *JASIST* | 3928 |
| *Library & Information Science Research* | *LISR* | 783 |
| *Library Quarterly* | *LQ* | 502 |
| *Online Information Review* | *OIR* | 1684 |
| *Scientometrics* | *SCIM* | 5926 |
| *Electronic Library* | *TEL* | 1821 |

**Table 4.** Classification system of research methods in LIS discipline(Chu&Ke,2017)

| Method | Definition |
|---|---|
| Bibliometrics | Bibliometrics is a method used for collecting publication and citation data. |
| Content analysis | Content analysis refers to collecting data by conducting systematic examination of texts or other passages in the contexts of their use. |
| Delphi study | The Delphi method is generally used for collecting data with a questionnaire from a group of experts to address a research problem in order to reach consensus and make forecasts via several rounds of exchanges. |
| Ethnography/field study | Ethnography and field study share many characteristics in data collection. Both can be applied when collecting data using multiple techniques, such as observation and interview, in a natural setting where participants live or work. |
| Experiment | Experiment is an established method for collecting data by following a procedure to test what is studied in either a laboratory or field setting, corresponding to laboratory experiments and field experiments described in(Palvia et al., 2007) list of research methods. |
| Focus groups | As a research method, focus groups refer to data collection via discussion of a research problem between a moderator and a group of participants. |
| Historical method | Historical method refers to collecting data by examining, synthesizing, summarizing, and interpreting existing published and unpublished materials related to a historical research problem. |
| Interview | Interview is a data collection technique where individual participants are asked questions relating to a research problem. |
| Observation | Observation is a method for gathering data via carefully and attentively watching and making notes on the subject being studied. |
| Questionnaire | Questionnaire, often known as survey, is a technique for data collection using a predefined list of questions. |
| Research diary/journal | Research diary or journal is a technique used to gather data about events, activities, thoughts, reflections, or other aspects by an individual who keeps the diary over a period of time. |
| Theoretical approach | Theoretical approach, as a research method, is a technique for gathering data through conceptual analysis, theoretical examination, or similar activities. |
| Think aloud protocol | Think aloud protocol is a research method intended to collect data about participants' cognitive activities via the verbal reports of their thoughts, called think alouds, while taking part in an experiment or performing some task. |
| Transaction log analysis | Transaction log analysis, as a research method, gains momentum when computerized systems are used for information processing and access. |
| Webometrics | Webometrics is defined as bibliometrics in the web environment, where webpages and websites are generally regarded as publications; with inlinks (i.e., links a webpage or site receives) being considered as citations and outlinks (i.e., links a webpage or site makes to others) being considered as references. |
| Other methods | Research methods other than the 15 mentioned above. |

**Classification system of research methods for academic papers in the field of LIS.** Regarding the framework for research methods in the LIS field, the mainstream classification systems currently used in research primarily include two approaches. The first is the classification system proposed by Järvelin and Vakkari (1990). These scholars initially introduced a framework encompassing research strategies and methods, encoding data collection methods in academic papers from a methodological perspective. This system has been consistently updated in subsequent studies, though its core content remains largely unchanged (Järvelin & Vakkari, 1990; Järvelin & Vakkari, 1993; Järvelin & Vakkari, 2021). The second is the classification system proposed by Chu and Ke (2017), which focuses on data collection methods. Subsequent scholars have widely adopted this system in their studies (Lou et al., 2021; Z. Zhang et al., 2021; C. Zhang et al., 2023). By analyzing articles published in three prominent LIS journals—JASIST, LISR, and JOD—they developed a classification system comprising 16 data collection methods. Considering factors such as the granularity of the classification systems and their historical development, this study adopts the methodological framework proposed by Chu and Ke (2017) to identify research methods in the corpus of academic papers. The specific classification system is detailed in Table 4. This system can assist LIS researchers in selecting appropriate methods by considering the applicability of each method to specific research questions. It is worth noting that, from a data collection perspective, the category of Bibliometrics in this system encompasses citation analysis, informetrics, and scientometrics.

**Selection of the classification model for research methods in LIS academic papers.** Previous studies have primarily relied on manual coding to identify research methods in academic papers, a process that is both time-consuming and labor-intensive, while also heavily dependent on expert knowledge (Chu & Ke, 2017; Järvelin & Vakkari, 1993). Given the substantial scale of the full-text corpus of LIS academic papers constructed in this study, an automated approach to research method classification is employed to identify the primary research methods at the document level for each paper. Inspired by the CogLTX model designed by Ding et al. (2020), Zhang & Tian, (2023) adapted this model for the task of research method classification, developing the CogFT (Cognize Full Text) model. This model demonstrates superior performance compared to traditional deep learning models based on pre-trained language models. Specifically, the CogFT model effectively extracts full-text features of academic papers while mitigating the noise introduced by irrelevant descriptions of research methods. Consequently, this study adopts the CogFT model for the task of document-level research method identification. CogFT is primarily designed for full-text encoding and method classification. First, due to the extensive length of academic papers and the presence of substantial noise unrelated to research methods, the full text is segmented into multiple chunks. Each input text chunk is then embedded using SciBERT (Beltagy et al., 2019) to represent it as a 768 dimensional vector. Subsequently, Self-Attention is employed to capture relationships between different text chunks and predict the probability that each input chunk constitutes a method summary. The four text chunks with the highest probabilities are concatenated with the first four chunks of the full text to form the research method summary of the paper. This summary is then fed as input to the research method classification network. The summary is processed by a standard BERT classifier, which assigns one of 16 predefined method labels. The training corpus for this model is derived from the dataset of English journal literature in the information science domain annotated by Chu and Ke (2017). The 16 categories of research methods defined by Chu and Ke (2017) are directly adopted as the model's classification labels. Ultimately, CogFT achieves an F1 score > 0.85, outperforming existing method annotation models.

Since a single paper may employ multiple research methods, the total number of identified methods exceeds the number of academic papers. Using the CogFT model to automatically classify research methods in the full-text corpus, the study ultimately obtains the classification results. The final classification yielded 31,401 distinct methodological instances drawn from 26,677 articles. Notably, 3,074 articles were found to incorporate multiple research methods. As illustrated in Figure 2, the top five research methods used in the papers are bibliometrics, experiment, questionnaire, theoretical approach, and content analysis, collectively accounting for over 75% of the total methods identified.

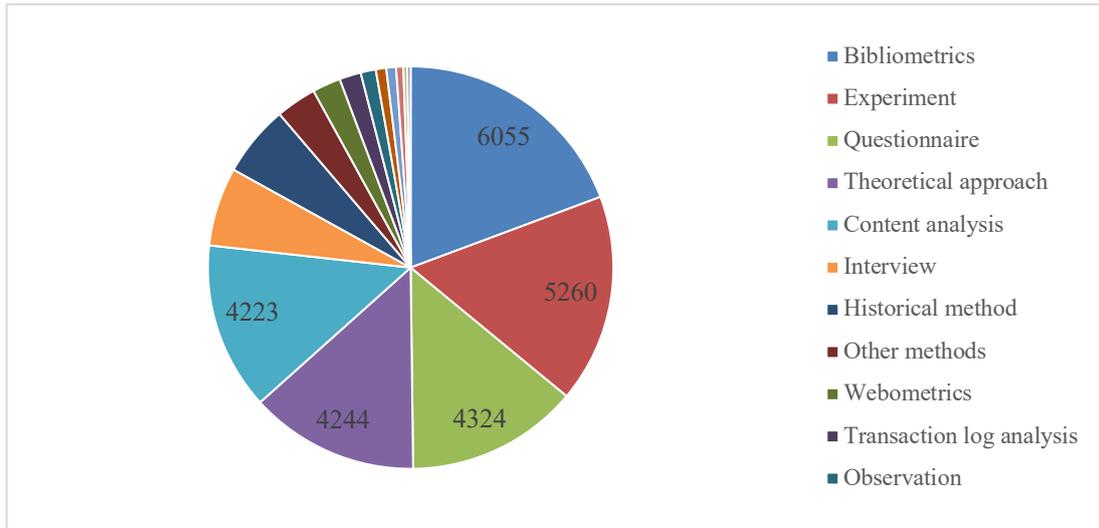

**Figure 2.** Classification results of research methods based on academic papers

### 3.4 Research Topic Clustering in LIS Academic Papers

This study employs the Top2Vec model (Angelov, 2020) to analyze academic paper abstracts and extract research topics. Top2Vec is an unsupervised topic modeling method based on deep learning. It automatically identifies latent topics from large-scale textual data and generates semantic vector representations of documents. The number of topic clusters was ultimately determined as 20 based on Topic Coherence (C_v score) calculations. This determination was subsequently supplemented by a manual review and summarization of high-frequency keywords under each topic. The resulting topic classification and nomenclature are presented in Figure 3. The most frequently used research topic among scholars in the papers is Scientometrics, covering 2,519 articles. This is followed by Information Theory, Information Retrieval, Text Mining, and Library Services, among others.

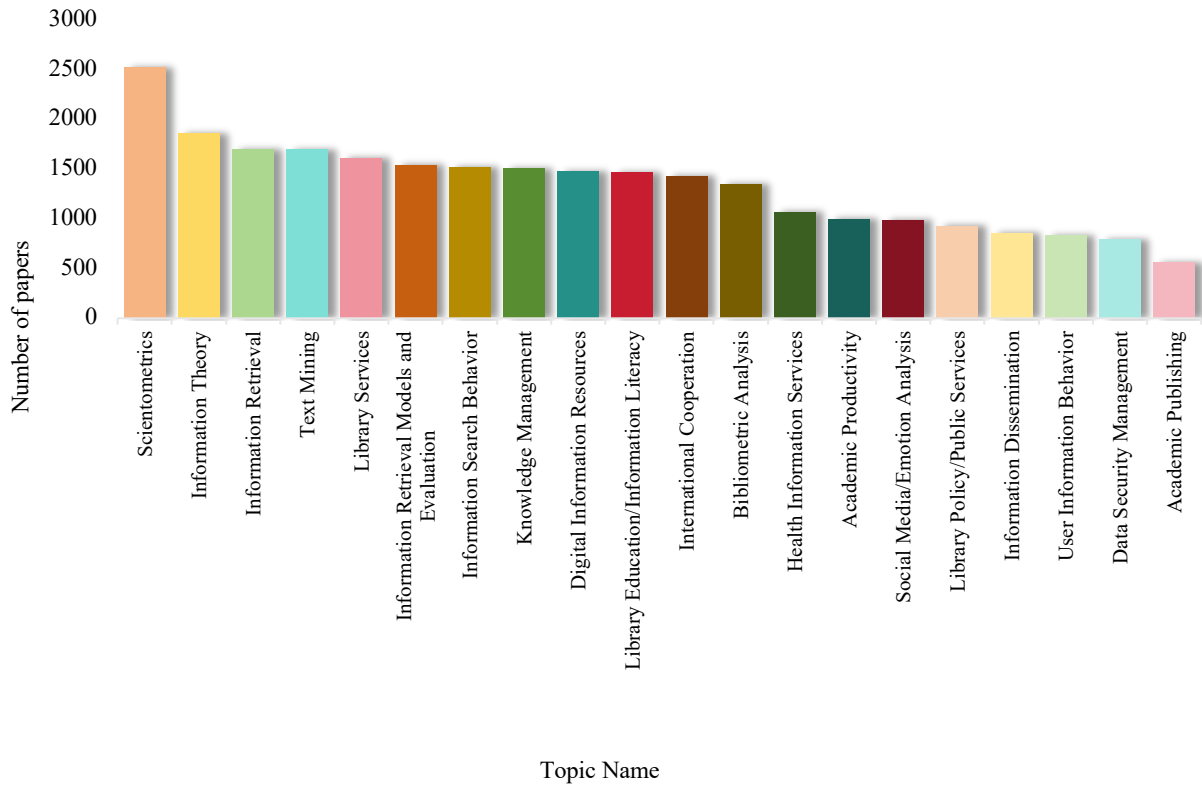

**Figure 3.** Classification results of research topics based on academic papers

## 3.5 Data processing for scholars' academic careers in the LIS field

This study investigates the evolution of research method selection among scholars in a specific field at different stages of their academic careers. In addition to the research methods identified earlier, it is necessary to perform author name disambiguation, calculate scholars' academic age. Based on these steps, we will select a subset of scholars to explore the evolution of their method choices throughout their careers.

**Scholar name disambiguation.** To examine the evolution of research method selection in scholars' academic careers, complete and accurate personal information is essential. This study utilizes OpenAlex (https://openalex.org/) to accomplish the task of author name disambiguation. OpenAlex is a free, open-access, large-scale scholarly resource indexing database that provides unique identifiers for various academic entities, including publications, authors, and institutions. It also offers multiple user-friendly API access methods. Among these, publication information can be retrieved using DOIs. Therefore, this study uses the DOIs from the metadata of academic papers to query OpenAlex, obtaining corresponding publication information and the unique identifiers of authors associated with each paper. These identifiers are then recorded and compared. Through this process, the study achieves accurate author name disambiguation results.

**Calculation of academic age of scholars.** To standardize the measurement of academic careers, this study defines a scholar's academic age as the time elapsed since their first publication. After completing the author name disambiguation process, the earliest publication of each author is retrieved from OpenAlex using their name. The publication year of this first paper is then extracted and used as the starting point for calculating academic age. Based on this starting point, the academic age of a scholar at the time of publishing a subsequent paper is calculated by taking the difference between the publication year of the paper and the year of their first publication, then adding one. The formula for calculating academic age is as follows:

$$AAS = PYA - EPY + 1 \qquad (1)$$

Where, $AAS$ stands for Academic Age of Scholar, $PYA$ stands for Publication Year of Article, and $EPY$ signifies the Earliest Publication Year. It is important to note that a scholar's academic age does not necessarily correspond to a specific range in their actual chronological age, as the real age at which scholars publish their first paper may vary. Therefore, this study employs academic age as the metric for investigating the use of research methods throughout scholars' academic careers.

**Criteria for selecting research method data in LIS scholars' academic careers.** After the above processing steps, the author has obtained the data of scholars' papers. Next, we will select scholars and summarize the relevant data of the papers they published during their research careers.

First, this paper filters papers based on the scholar's authorship information. In the field of LIS, it is generally accepted that when a scholar is listed as the first author or the last author (typically the corresponding author), it often indicates a significant contribution to the paper (Frandsen & Nicolaisen, 2010; C. Zhang et al., 2025). Therefore, considering collaboration and division of labor among scholars, this study selects papers where the scholar is either the first author or the corresponding author and identifies the research methods used in these papers as those employed by the scholar during their academic career stage. In the corpus of academic papers used in this study, 14,856 articles have the same individual as the first author and corresponding author, while 8,471 articles have different individuals in these roles. Accordingly, when counting authors, this study considers both the first author and corresponding author for articles where these roles are distinct. For articles where the first author and corresponding author are the same, the author is counted as a single individual.

Second, to ensure the validity and reliability of the data, it is necessary to remove outliers in scholars' academic age. The Interquartile Range (IQR) method, which is based on the quantiles of the data, is effective in excluding extreme values and is not influenced by outliers. Therefore, this study employs the IQR method to identify and remove outliers in academic age. The academic age data of the scholars were first sorted from smallest to largest. Formula (2) calculates the inter - quartile range. $Q1$ represents the lower quartile, which is the value at the 25th percentile. $Q3$ represents the upper quartile, which is the value at the 75th percentile. Second, values in the academic - age data that are less than the lower limit or greater than the upper limit may be regarded as outliers. Formula (3) and Formula (4) calculate the upper and lower limits of the academic age respectively.

Finally, the calculated outliers of the authors' academic ages are eliminated, and a total of 14,622 authors' data are obtained. Figure 4 shows the distribution of authors' academic age. Authors with an academic age of 1 form the largest group.

$$IQR = Q3 - Q1 \qquad (2)$$
$$Upper = Q3 + 1.5 * IQR \qquad (3)$$
$$Lower = Q1 - 1.5 * IQR \qquad (4)$$

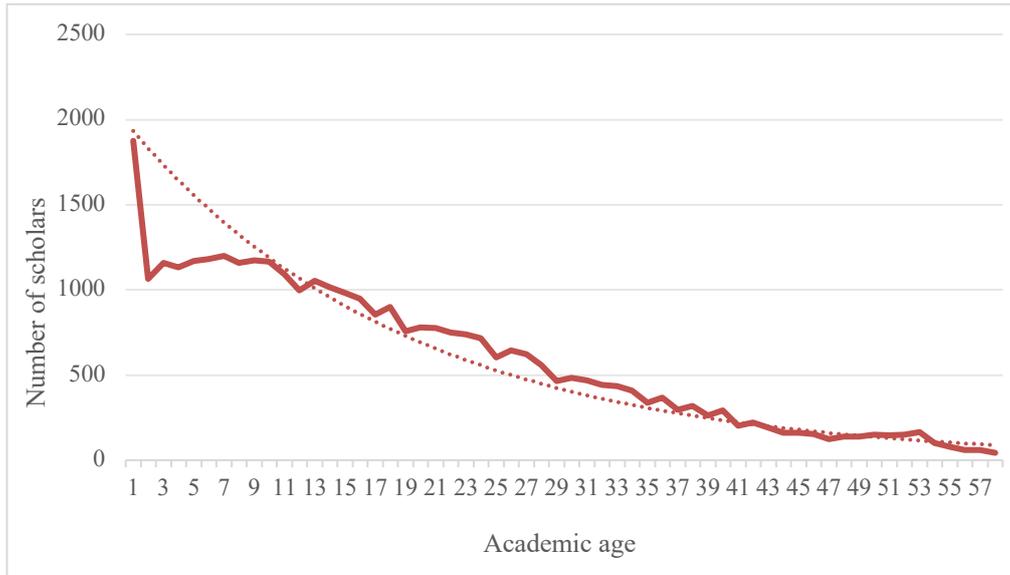

**Figure 4.** Distribution of authors' academic age

To analyze the relationship between academic age and method use, we establish a classification. We set the maximum academic age at 61 years (95th percentile). Following prior research (Chowdhary et al., 2024), we define three groups: young scholars (academic age < 7), middle-aged scholars (7–14), and senior scholars (>14).

Finally, the selection of scholars was conducted. Given that method evolution varies across individual academic careers, this paper focuses on scholars with longer, consistently active research careers to capture the overall trend in method choice patterns among the majority. Therefore, senior scholars with an academic age greater than 14 years and a consistent publication record at five-year intervals were selected. This resulted in a cohort of 435 senior scholars, encompassing 6,116 published articles.

## 4  Results

### 4.1  Correlation analysis of academic age and research methods of scholars in the field of LIS

In this section, we address *RQ1* by exploring whether differences exist in the selection of research methods among scholars at different academic ages. To achieve this, statistical methods for difference analysis and testing are applied. Common methods for difference analysis include the t-test, analysis of variance (ANOVA), and the chi-square test. The chi-square test is suitable for scenarios where both independent and dependent variables are categorical. Therefore, this section employs the chi-square test to measure the frequency differences in method selection among scholars belonging to three distinct academic age groups. Each research method is independently subjected to a chi-square test. Figure 5 presents the results of the chi-square statistics.

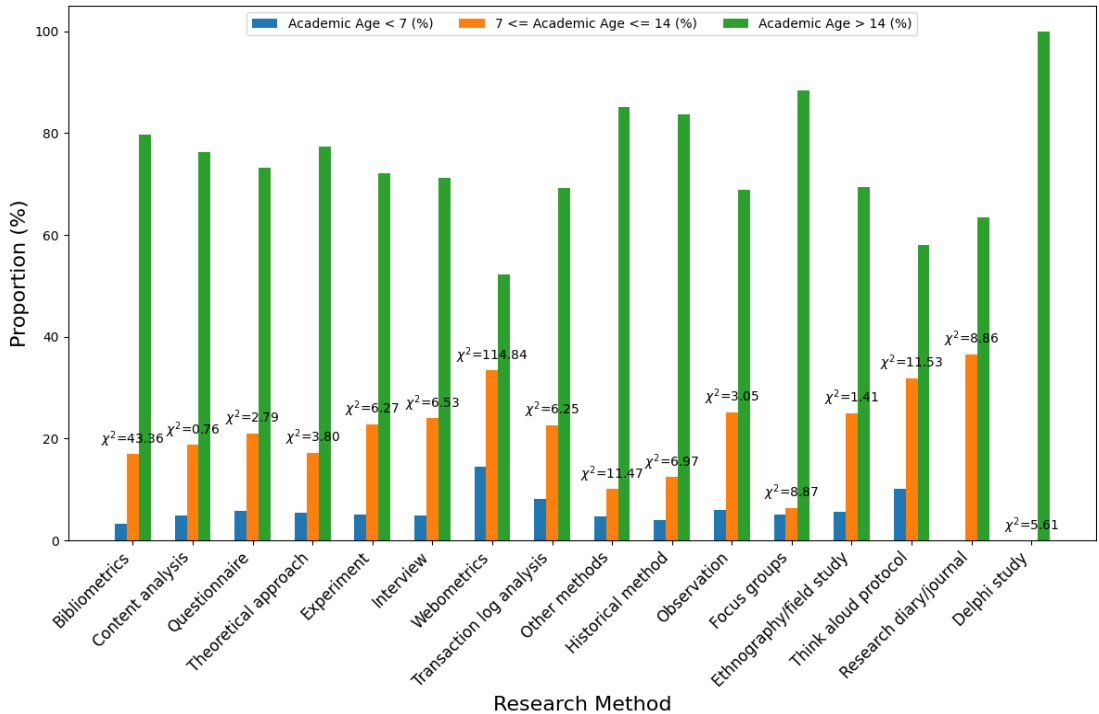

**Figure 5.** Statistical differences in the frequency of method selection among scholars in different academic age groups

Within the specific field, the usage proportions of different research methods exhibit significant variation. Among these, bibliometrics has the highest proportion at 30.02%, indicating that this method is the most commonly employed by scholars in the field. In contrast, focus groups, ethnography/field study, think aloud protocol, research diary/journal, and delphi study are used very infrequently, each accounting for less than 1% of the total. This suggests that these methods are rarely adopted in research. Out of the 16 research methods examined, only 6 show no significant differences in selection frequency across academic age groups. This indicates that scholars at different stages of their academic careers exhibit distinct preferences in their choice of research methods. When scholars are in the early stage of their academic careers, that is, when their academic age is less than 7, there are 3 methods they tend to choose. When scholars' academic age is between 7 and 14, there are 6 methods they prefer. When scholars' academic age is greater than 14, there are 4 methods they are inclined to select. Obviously, scholars in their middle - aged period tend to choose a larger variety of methods. In addition, this paper uses the chi - square value to judge the degree of significance of differences in method selection at different stages of the academic career. The top three methods with the largest chi-square values are webometrics (Chi2=114.8354***), bibliometrics (Chi2=43.3623***) and think aloud protocol ( Chi2=11.5278**). Webometrics and bibliometrics are the methods preferred by academics in their younger and middle-aged years. Think aloud protocol is the method preferred by academics in their senior years.

**4.2   Differences in research methods used by scholars of different academic ages in different periods**

To further examine how method usage varies across time periods and academic ages, we first analyze the diversity of methods used by scholars at different career stages over time. We then identify the top five methods per age group and explore trends in their usage frequency over publication years.

**Types of research methods used by scholars in different academic age groups across publication periods.** To investigate the diversity of research methods used by scholars at different career stages over time, this study constructs a heatmap based on five-year intervals of publication years and academic age groups. Since the number of publications varies across time periods, the data on the types of research methods used are normalized to ensure comparability.

As shown in Figure 6, the darker regions are predominantly concentrated in the period from 2000 to 2024 and among scholars with academic ages ranging from 1 to 50 years. This indicates that since 2000, scholars across various academic age groups have increasingly adopted a more diverse range of research methods. Furthermore, for each publication period after 2000, the number of research methods used initially increases and then decreases as scholars progress in their academic age. During the early period of 1990–1994, most academic age groups are represented by light green or light yellow hues. This suggests that, regardless of academic age, scholars during this time employed a relatively limited variety of research methods. From 2000 to 2014, the colors gradually deepen, particularly among scholars in the 11–45 academic age range, where the values reach as high as 0.93 or even 1. This indicates that scholars in this range utilized nearly all available types of research methods, reflecting a significant diversification in their methodological approaches. In the period of 2015–2024, the color distribution shifts again, with the hues for the 31–50 academic age group becoming lighter. The trend for the 11–30 academic age group shows that these scholars maintained a high diversity in research method usage over an extended period, likely due to their being in the prime of their academic careers, where they possess the capability and resources to experiment with a wide range of methodologies. For scholars in the 46+ academic age group, the overall number of research methods used is relatively low.

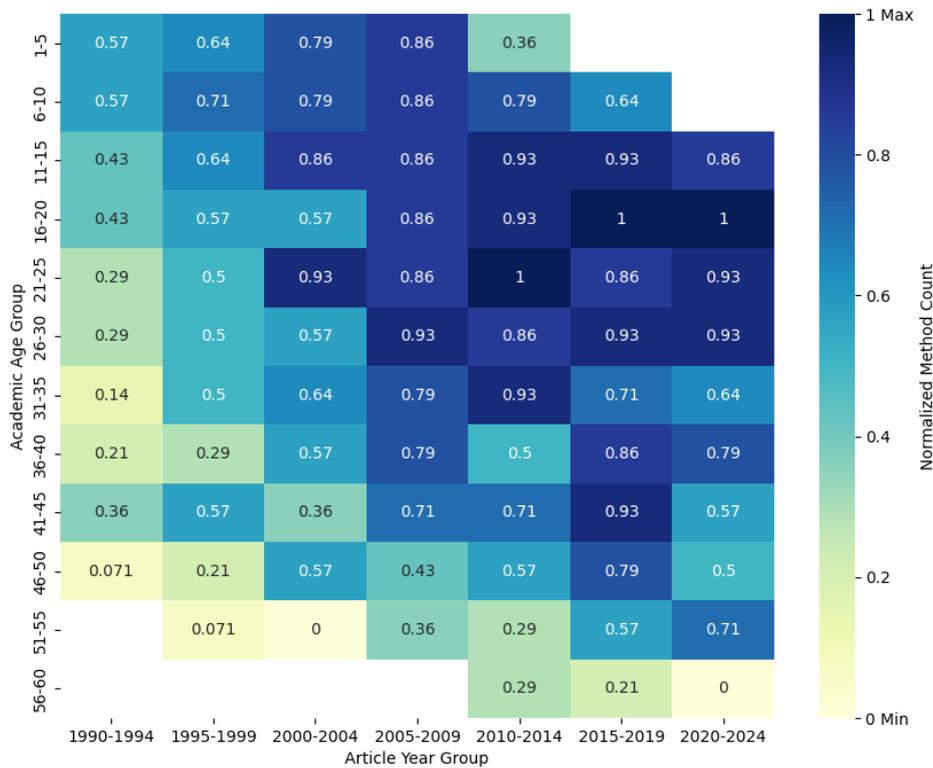

**Figure 6.** Heat map of the types of research methods used

As shown in Figure 6, the diversity of research methods used exhibits dynamic changes across different academic age groups and publication periods. Over time, there is an overall trend toward increased method diversity, though the extent and timing of these changes vary among academic age groups. The middle - aged academic age group has maintained a high level of research method diversity over a long period. Young scholars are continuously increasing the number of types of research methods they use, while senior scholars remain relatively stable.

**Top five research methods used by scholars of different academic age groups.** In order to deeply analyse which types of research methods are more popular among scholars at different stages of their academic careers, this paper summarizes the annual percentage of the top five research methods used by different academic age groups. The specific situation is shown in Figure 7.

As depicted in Figure 7, the most frequently used research methods among scholars remain relatively consistent across different academic age groups. For senior scholars, bibliometrics consistently ranks first in usage, with its proportion showing an upward trend—rising from 19.61% to 31.81%. This indicates that bibliometrics is highly favored by scholars. It also underscores the dominant role of bibliometrics in the field of information science. Questionnaire and content analysis maintain stable usage proportions across all academic age groups, consistently ranking second and third, respectively. This reflects the broad applicability and enduring demand for these methods. Theoretical approach also persists throughout scholars' academic careers, highlighting the guiding role of theoretical research in academic inquiry. Webometrics ranks fifth in usage among younger scholars, indicating its popularity within this group. Meanwhile, experiment exhibits relatively stable usage proportions among mid-career and senior scholars, ranking fourth and fifth, respectively. This suggests that experiment becomes an important research tool as scholars accumulate experience and enhance their research capabilities.

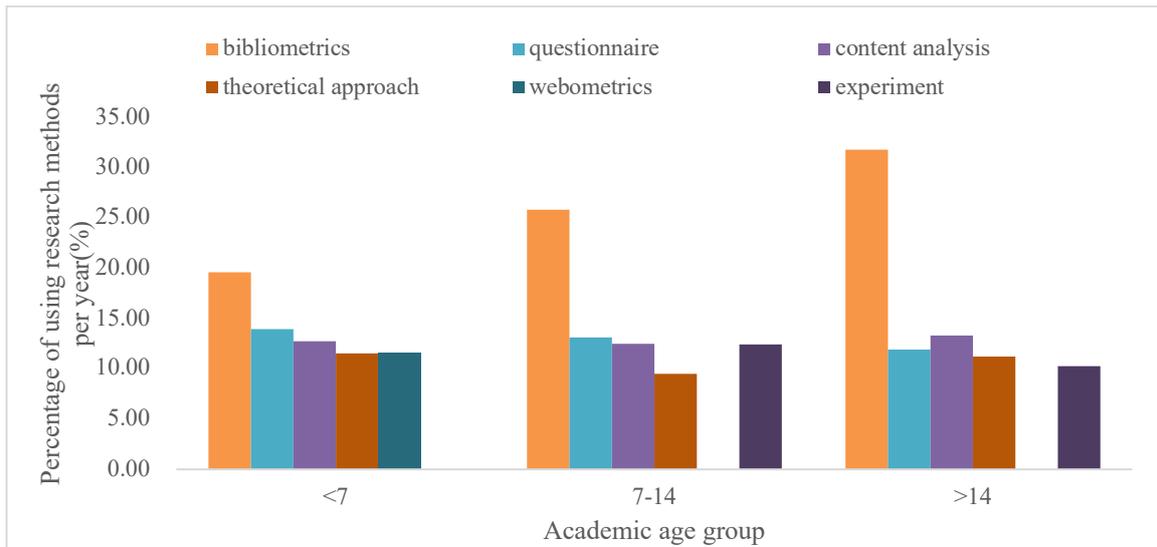

**Figure 7.** Top five research methods by usage proportion across different academic age groups

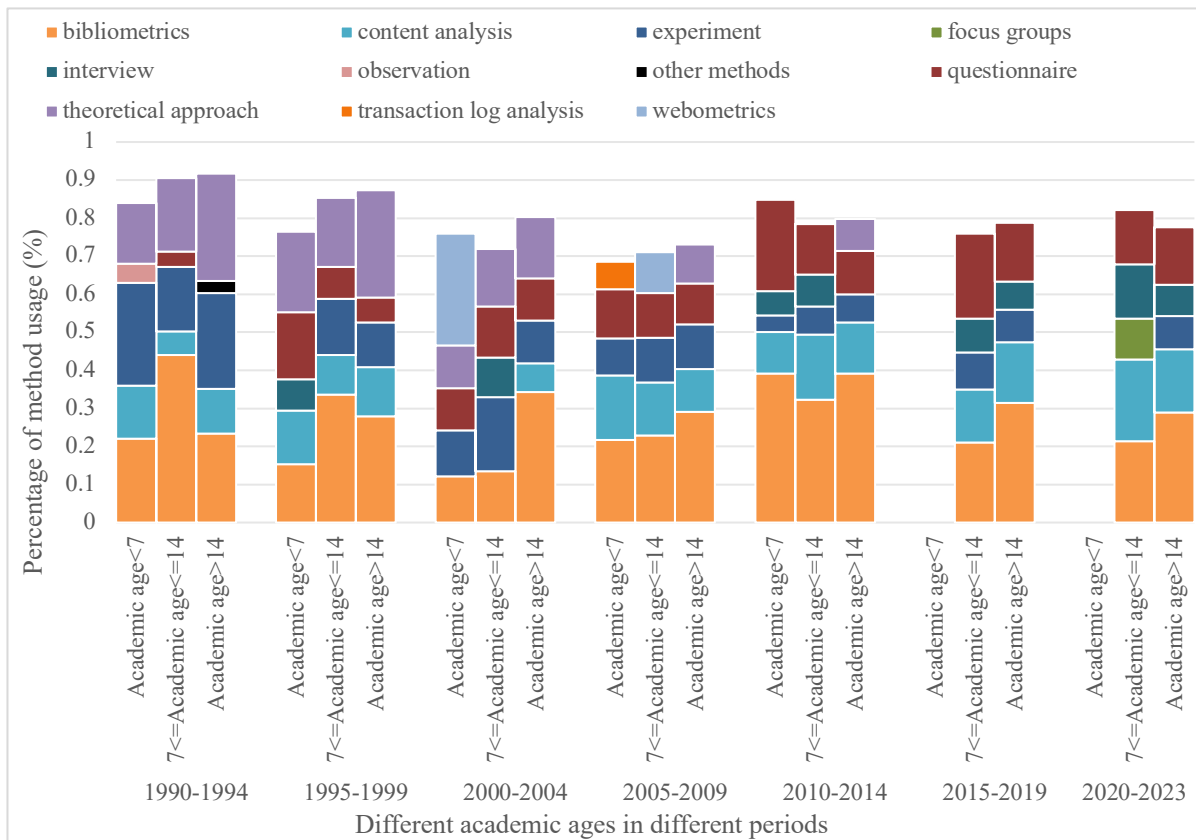

**Figure 8.** Top five research methods used by different academic age groups. An academic age of < 7 years indicates young scholars, 7–14 years indicates middle-aged scholars, and > 14 years indicates senior scholars.

The usage proportions of research methods among scholars in different academic age groups also vary over time. As shown in Figure 8, the trend of the top five research methods in terms of percentage of use varies slightly across different academic ages in different periods of time. Bibliometrics covers the range of academic careers of scholars in all periods of time and is consistently high in terms of percentage of use. It is followed by content analysis, experiment and questionnaire. This confirms the trend of the overall top five used research methods as reflected in Figure 7. Since 2000, webometrics has been highly favored by young scholars, and it ranked fifth among the methods used by middle - aged scholars from 2005 to 2009. This may be attributed to the fact that young scholars from 2000–2004, as they advanced in age and experience, transitioning into middle - aged scholars, retained their preference for bibliometrics. For theoretical approach, the method was highly preferred by scholars at all academic career stages from 1990-2000, with a share of around 20%. However, its ranking gradually declined after 2000 and disappeared from the top five list after 2015. This shift may be linked to the rise of emerging technologies, such as machine learning models, which have increasingly been applied in academic papers, potentially displacing other traditional methods. Certain methods, such as transaction log analysis and focus groups, appear prominently only in specific periods and academic age groups.

Overall, scholars in different academic age groups exhibit variations in their use of research methods across different time periods. Over time, the usage proportions of certain methods, such as bibliometrics and content analysis, have gradually increased across all academic age groups. In contrast, the usage proportions of more traditional methods, such as interview and theoretical approach, have declined.

**Evolution of research method usage among scholars at different career stages.** This section primarily focuses on the changing trends and fluctuation characteristics of each research method over time across different academic career stages, rather than directly comparing the absolute usage volumes between different methods. Therefore, using original frequency data instead of normalized data allows for a more intuitive representation of the true fluctuation amplitude of each method's trend line. Figure 9 presents the evolving trends in the frequency of usage for the 16 research methods among scholars in different academic age groups. Overall, the usage frequency of most methods shows significant fluctuations between 1990 and 2020. This indicates that scholars' adoption of research methods has not been stable over the years. These fluctuations underscore the diversity and dynamism of research methodologies in academic inquiry. Moreover, for each research method, the trends in usage frequency appear consistent across the three academic career stages. This may be attributed to the inherent characteristics of the methods themselves, where a method gaining popularity in a particular period leads to its widespread adoption by scholars across all age groups.

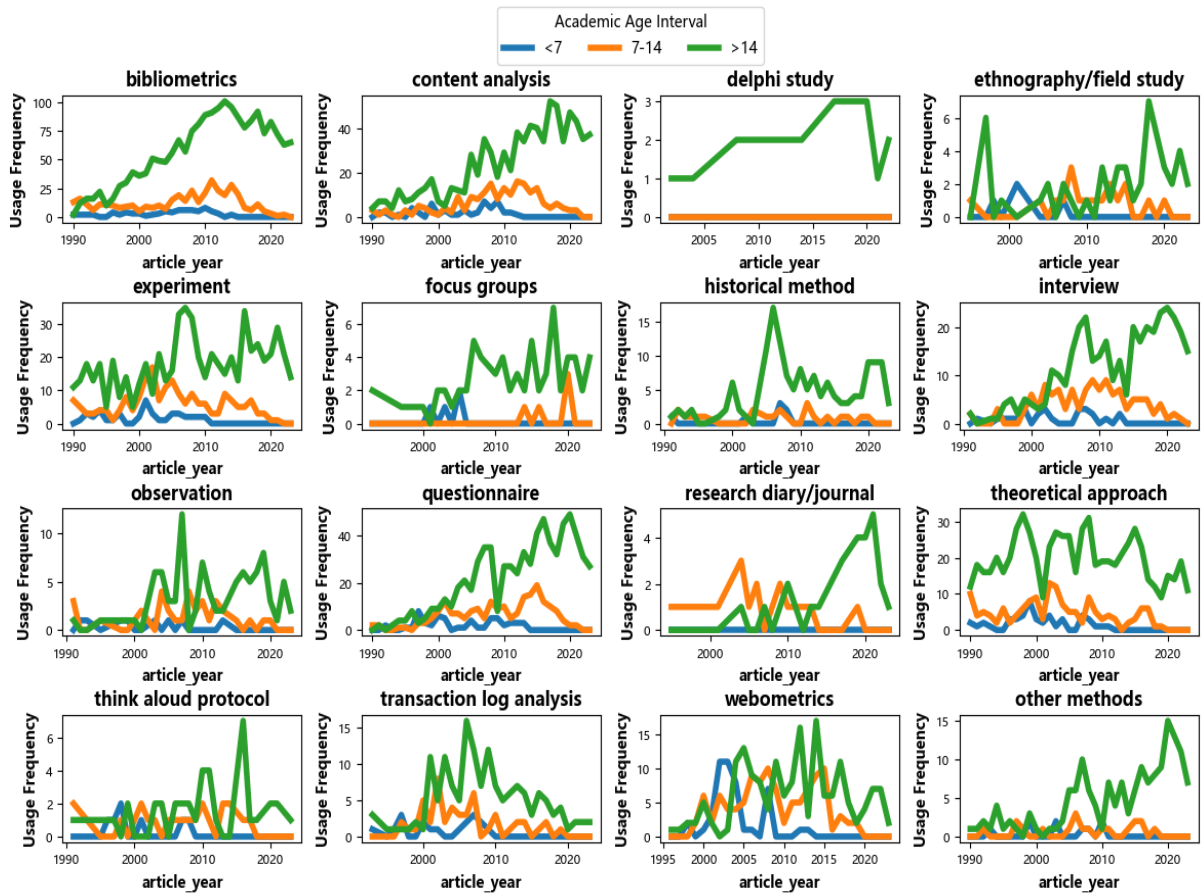

**Figure 9.** Evolutionary trends in the use of different research methods by scholars at different stages of their academic careers

In papers published by scholars in the senior stage of their careers, the use of methods such as bibliometrics, content analysis, interview, and questionnaire exhibits a pronounced upward trend. Notably, bibliometrics, which had relatively low usage frequency from 1990 to 1995, experienced rapid growth starting in 1995 and maintained high usage frequency between 2010 and 2020. This trend may be linked to the rapid development of scientometrics and the increasing emphasis on Bibliometric Analysis in academia. In contrast, the use of experiment and theoretical approach remains relatively stable, indicating that theoretical research continues to hold a significant position in academic inquiry. During the period of 2005–2010, methods such as experiment, historical method, interview, observation, and transaction log analysis reached a notable peak in usage. This suggests that scholars during this five-year period were inclined to employ a diverse range of research methods rather than limiting themselves to commonly used or popular approaches.

Apart from the aforementioned methods, most other methods do not exhibit significant trends in usage frequency due to their inherently low adoption rates. When scholars are in the early stages of their careers, the use of webometrics shows a leading trend. This can be attributed to the influence of internet technology on academic research methods, as well as the greater willingness of younger scholars to adopt and apply emerging technologies. When scholars are in their middle age, the frequency of using all kinds of research methods increases compared with that in their younger age. This may be because scholars' careers are relatively stable in middle age and the valuation risk is relatively reduced. Therefore, scholars will try to use a variety of research methods to achieve self - breakthroughs and enhance their academic influence.

As shown in Figure 9, the trends in the usage frequency of different research methods between 1990 and 2020 vary significantly. Emerging methods, such as webometrics, exhibit rapid growth trends driven by technological advancements. In contrast, more traditional methods, such as questionnaire and theoretical approach, maintain relatively stable usage frequencies. The adoption of research methods is influenced by a variety of factors, including disciplinary developments, technological progress, and shifts in research hotspots. Scholars adapt their methodological choices over time to align with the practical demands of their research.

### 4.3  Differences in Research Method Usage Among Scholars of Different Academic Age Groups Across Various Research Topics

This section examines the diversity of research methods employed by scholars at different career stages from the perspective of research topics. First, it analyzes the prevalent research themes among scholars in various academic age groups. Subsequently, it investigates the combined preferences of these scholars regarding both methods and topics. Furthermore, the study explores differences in method usage across research topics stratified by academic age. Finally, it traces the evolving trends in the frequency of research topic usage among scholars at different career stages based on publication year.

**Top Five Research Topics Among Scholars of Different Academic Age Groups.** To further investigate which types of research topics are favored by scholars at different career stages, this study aggregates the annual proportion of the top five research topics utilized by various academic age groups. The detailed results are presented in Figure 10. The most frequently used research topics remain largely consistent across academic age groups. Four topics—International Cooperation, Scientometrics, Bibliometric Analysis, and Information Search Behavior—recur among the top five topics consistently shared by all three academic age cohorts. In contrast, topics such as Library Services, Information Retrieval, and Information Theory each appear only once across the three groups.  For scholars at all career stages, Scientometrics consistently ranks as the most prevalent topic. International Cooperation shows the smallest proportional variation among the three groups, reflecting its universal relevance. Younger scholars are often driven by trends in global academic exchange, while middle-aged and senior scholars leverage established networks to advance collaborative research. This consistency suggests that motivation for International Cooperation persists throughout academic careers.

From the perspective of different academic age groups, early-career scholars tend to focus on algorithm application-oriented topics such as information retrieval. Mid-career scholars are more inclined to conduct empirical studies on service models and policy effects—exemplified by topics like library services. Senior scholars show greater engagement with theoretical advancements, particularly in areas such as information theory. This distribution reflects generational differences in research priorities.

Furthermore, the proportional emphasis on these topics among scholars of different academic ages has evolved over time. As shown in Figure 11, the trends in the top five research topics vary slightly across different periods and academic age groups. The topic of Scientometrics consistently appears throughout nearly all periods and maintains a high proportion of research focus. Prior to the 21st century, significant attention was devoted to topics such as information retrieval models and evaluation, as well as digital information resources. In contrast, research on topics like social media/sentiment analysis and Bibliometric Analysis has increased since the 2000s. From the perspective of academic age groups, young scholars are technology-sensitive and trend-responsive, leading in topics like Information Search Behavior and Information Retrieval. Middle-aged scholars exhibit strengths in method integration and domain expansion, effectively bridging traditional topics like Scientometrics with emerging technical topics such as text mining. They serve as a nexus between classical and emerging themes. Senior scholars focus on theoretical refinement and practical consolidation, maintaining engagement with classical domains like Scientometrics and Information Theory. In summary, young scholars drive technological innovation, middle-aged scholars integrate themes, and senior scholars provide theoretical depth, collectively facilitating intergenerational knowledge transmission and disciplinary advancement.

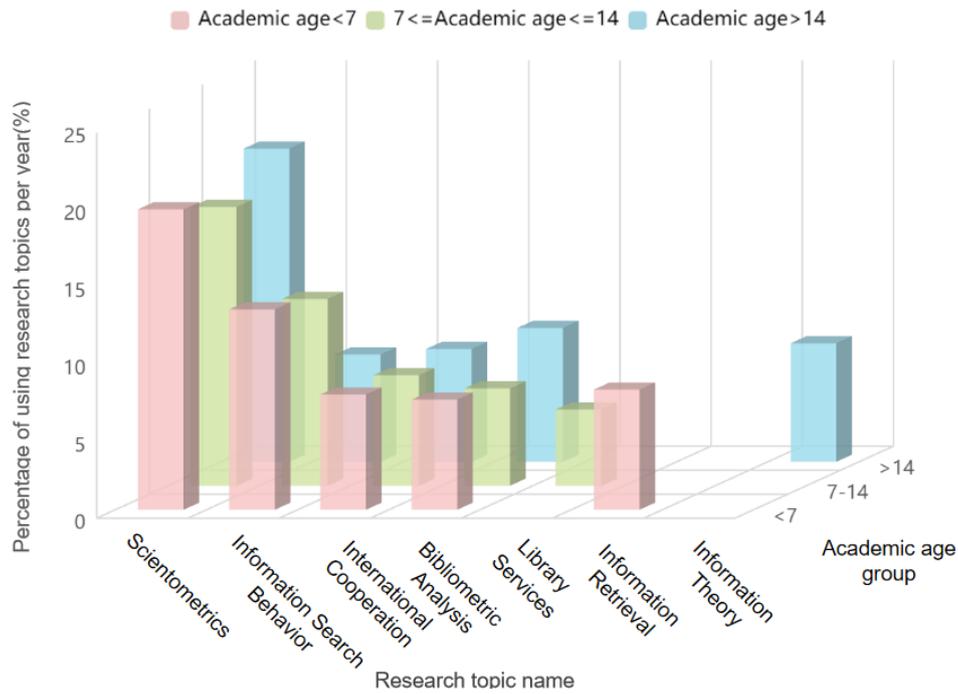

**Figure 10.** Top Five Research Topics by Usage Proportion Across Academic Age Groups

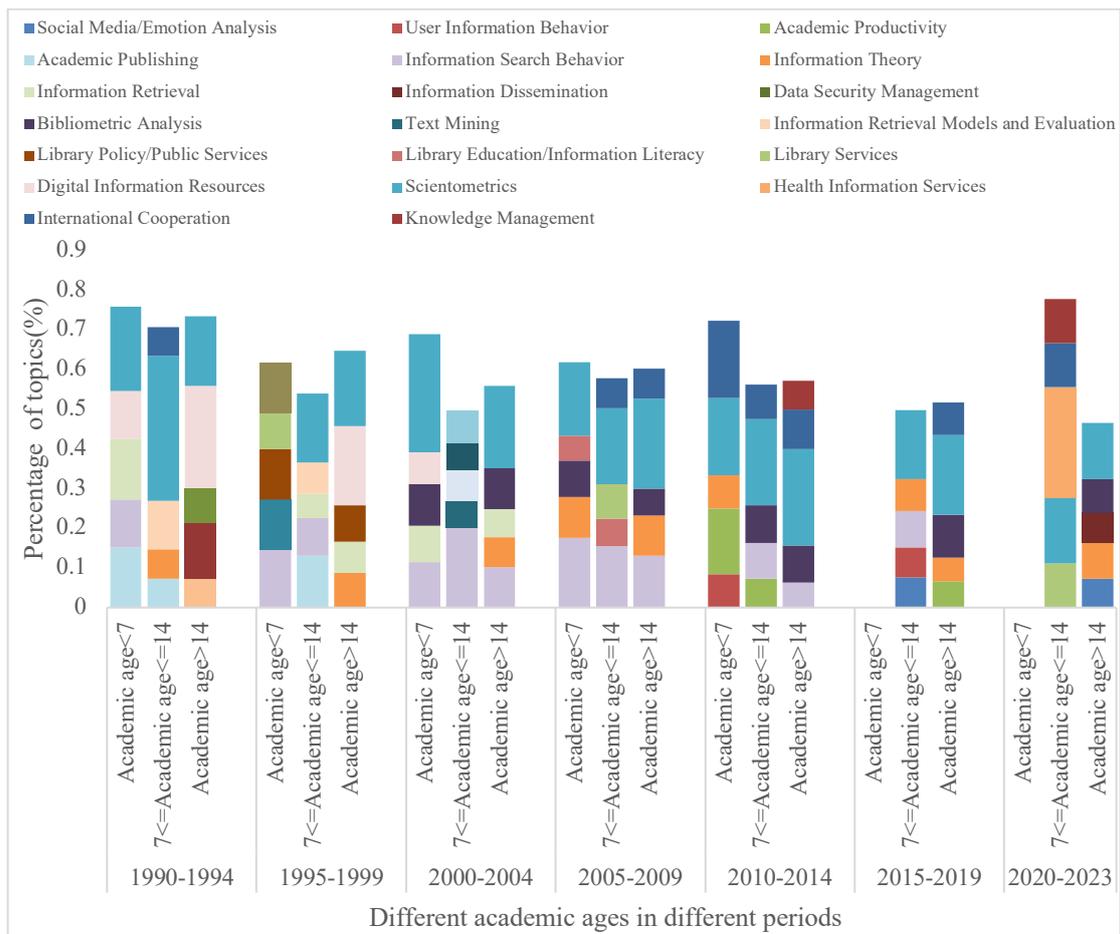

**Figure 11.** Top Five Research Topics by Usage Proportion Across Academic Age Groups. An academic age of < 7 years indicates young scholars, 7–14 years indicates middle-aged scholars, and > 14 years indicates senior scholars.

**Integrated Preferences of Research Methods and Topics Among Different Academic Age Groups.** To visualize the comprehensive preferences of scholars across various age groups regarding research methods and to pics, three three-dimensional diagrams were constructed. In these diagrams, each radial point represents an individual scholar within a specific academic age group, reflecting three integrated dimensions: the number of publications, the diversity of research methods used, and the variety of research topics covered. For better observation and interpretation, we also provide an interactive, rotatable annotated version.[1]

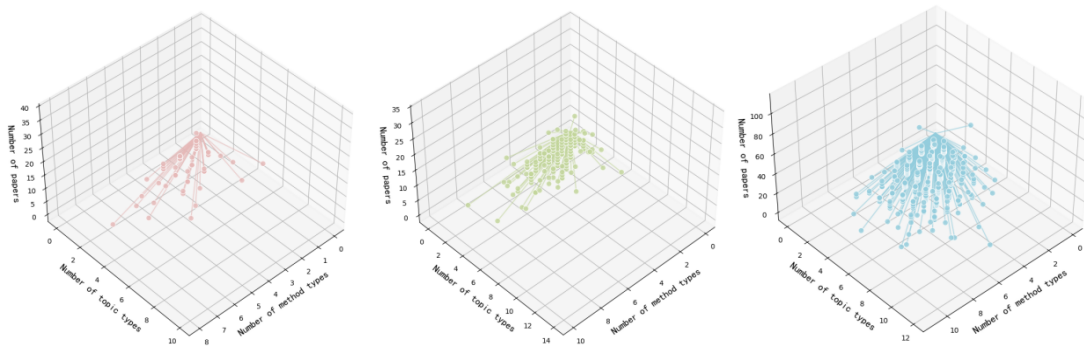

**Figure 12.** Three-Dimensional Representation of Publication Count, Research Methods, and Topic Diversity Across Scholar Groups

As shown in Figure 12, the three diagrams from left to right represent young, middle-aged, and senior scholars, respectively. Among young scholars (pink diagram), data points are predominantly clustered in the low-to-medium range of methodological diversity, while thematic diversity remains relatively concentrated. This suggests that young scholars often experiment with multiple methods while focusing deeply on a limited set of core topics—a pattern summarized as "multiple methods, limited themes." Middle-aged scholars (green diagram) exhibit broader distributions in both method and topic diversity. Their increased publication output reflects an ability to adapt familiar methods to a wider range of topics, demonstrating how they "leverage established methods to expand thematic coverage," thereby accelerating research productivity. Senior scholars (blue diagram) display a more balanced profile across all three dimensions. The evenly distributed and rounded shape of the point cloud indicates synergistic growth in method versatility, topic breadth, and sustained publication output, achieving what may be termed a "triangular balance" of method maturity, topic expansion, and consistent productivity.

Overall, the three groups illustrate a clear evolution: young scholars explore method tools, middle-aged scholars reuse and adapt these methods to expand thematic scope, and senior scholars ultimately achieve a harmonious balance among method, topic, and output.

**Usage of Research Methods Across Academic Age Groups within Different Research Topics.** This section presents three bubble heatmaps illustrating the relationship between research topics and method usage for scholars in each academic age group. In each visualization, the color of a bubble represents the logarithm of the usage frequency of a specific research method within a particular topic among the corresponding scholar group. Darker colors indicate higher method usage within the associated topic. For the convenience of observation, we also provide an interactive three-dimensional bubble chart.[2]

Figure 13 illustrates the usage of research methods by young scholars across different research topics. From a method-to-topic adaptation perspective, bibliometric methods are most frequently employed in the field of Scientometrics, indicated by a dark red hue, followed by webometric methods. This aligns with the core demand for quantitative analysis of academic literature in LIS research. Webometric methods are also widely applied in topics related to International Cooperation. In themes such as Information Search Behavior, methods including questionnaires, experiments, and interviews are more prevalent, reflecting the reliance on multi-method data collection in context-specific user behavior studies. Although content analysis is applied across a broad range of thematic areas, its general-purpose nature results in relatively moderate frequency within any single topic.

Figure 14 presents the method preferences of mid-career scholars across various research topics. In the domain of Scientometrics, bibliometric methods continue to demonstrate the highest frequency of use. Additionally, this approach is frequently applied in topics such as International Cooperation, Academic Productivity, and Bibliometric Analysis. From a vertical perspective, the theme of Information Search Behavior employs a variety of methods, including questionnaire, experiment, and transaction log analysis. Horizontally, methods such as theoretical analysis, questionnaire, and interview exhibit moderate frequency—indicated by orange-yellow hues—across multiple topics, including Information Retrieval, Information Theory, Health Information Services, and Library Services. This reflects the adaptability of these methods to diverse research themes. Overall, high-frequency methods are concentrated in data-driven strategies such as bibliometrics, experiment, and questionnaire, highlighting the emphasis on measurability in academic research.

---

[1] https://jiayihao-njust.github.io/evolution/Three-dimensional%20diagram%20of%20Young%20scholars.html
[2] https://jiayihao-njust.github.io/evolution/3d_bubble_chart.html

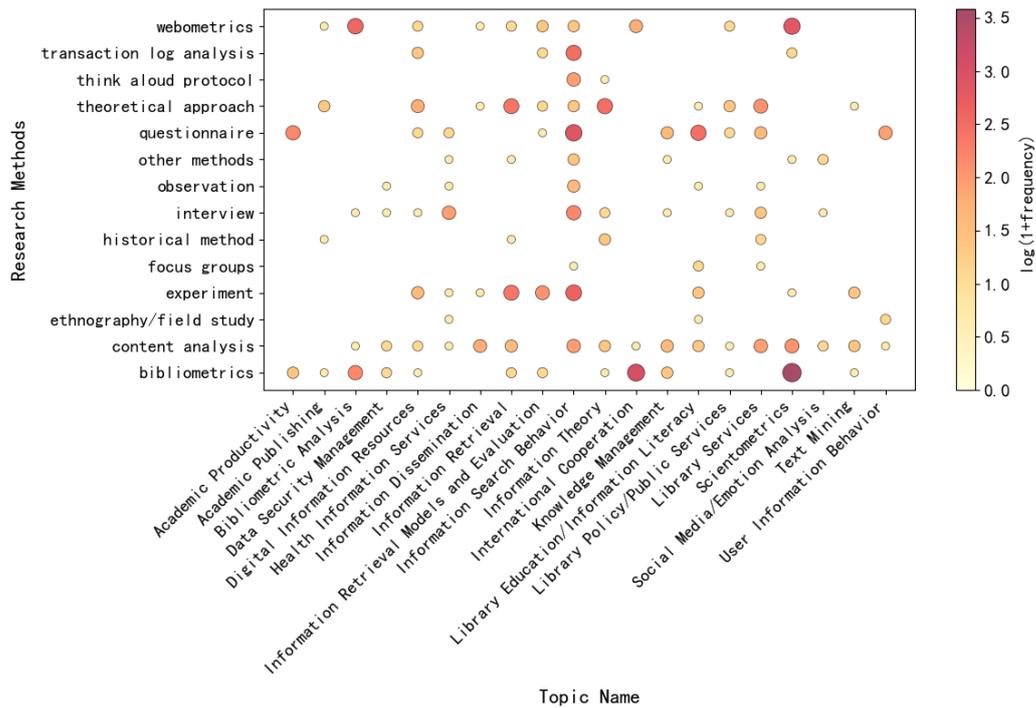

**Figure 13.** Heatmap of Research Method Usage by Young Scholars Across Research Topics

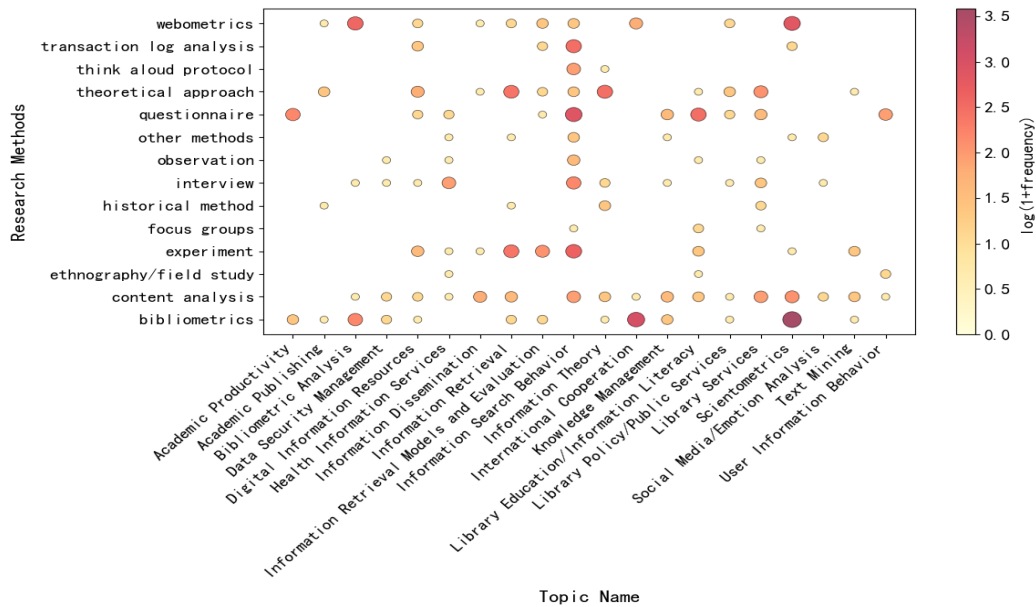

**Figure 14.** Heatmap of Research Method Usage by Middle-aged Scholars Across Research Topics

Figure 15 reveals the relationship between research methods and topics among senior scholars. Within the domain of Scientometrics, bibliometric methods are employed with significantly higher frequency compared to early- and mid-career groups. Both theoretical analysis and questionnaire methods demonstrate broad applicability across multiple research topics. Theoretical approaches are frequently applied in topics such as Information Theory, while questionnaires are commonly used in areas like User Information Behavior and Library Education/Information Literacy. This pattern highlights senior scholars' maturity in adapting methodologies to diverse thematic contexts. Additionally, content analysis achieves balanced penetration across multiple topics, indicating its versatile utility. This pattern of multifaceted method integration permeating research topics illustrates senior scholars' scholarly depth and their capacity to synthesize complex research approaches.

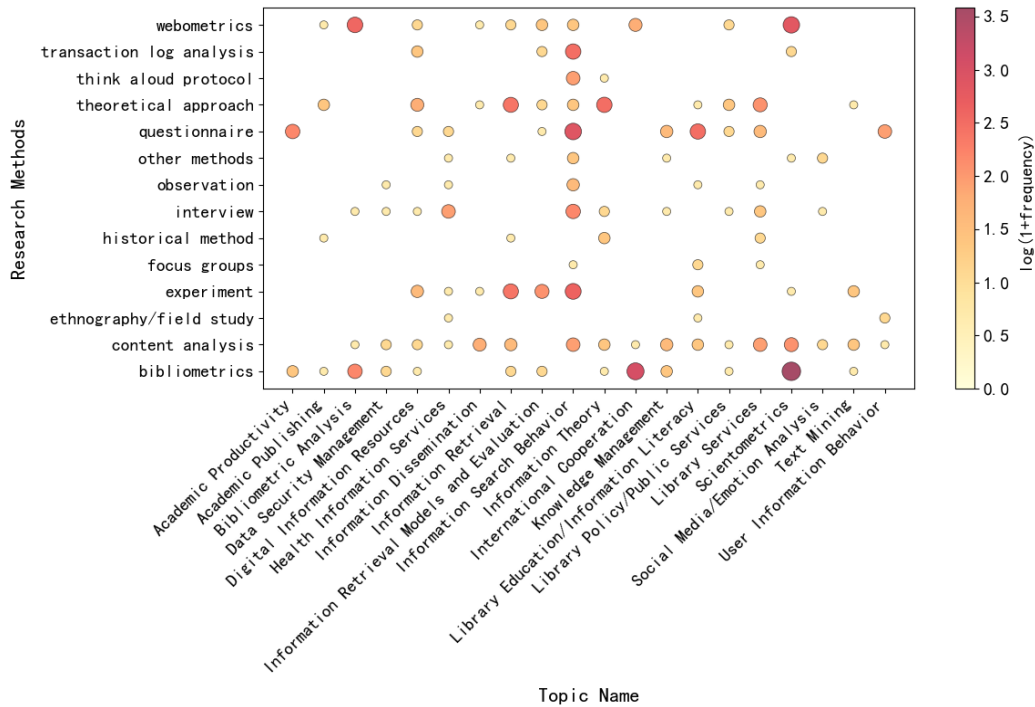

**Figure 15.** Heatmap of Research Method Usage by Senior Scholars Across Research Topics

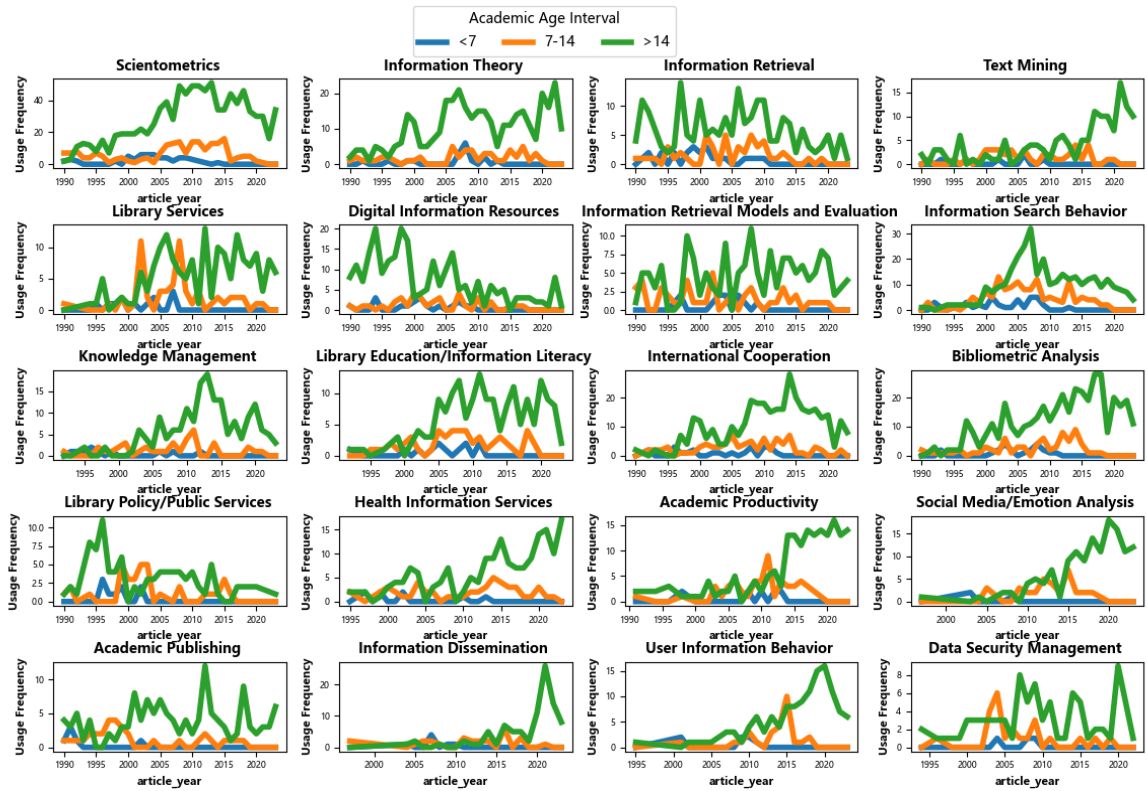

**Figure 16.** Trends in Research Topic Application

**Evolving Trends in Research Topic Usage Among Scholars at Different Career Stages.** Figure 16 illustrates the evolving trends in the usage frequency of the 20 research topics among scholars at different career stages between 1990 and 2023. Overall, all topics exhibit significant fluctuations in usage frequency, reflecting the dynamic nature of research interests. From a thematic perspective, the application frequency of classical quantitative topics such as Scientometrics and Academic Productivity has gradually increased over time. Topics like Digital Information Resources and Library Policy/Public Services were relatively prevalent before the 21st century but subsequently declined in prominence. The topic of Information Search Behavior peaked around 2010, particularly among senior scholars. This surge can be attributed both to the systematic development of theories and methods in the field, which attracted scholars across career stages, and to the leading role of experienced scholars who leveraged their expertise to advance and deepen the research agenda. Technology-driven topics exhibit a late-mover advantage: emerging themes such as Social Media/Sentiment Analysis and Text Mining saw a sharp increase in adoption among senior scholars after 2010. Their strong resource integration capabilities enabled breakthroughs in technically complex topics. From the perspective of scholarly behavior, senior scholars maintain a sustained authority in mature topics such as Scientometrics, while also demonstrating strong adaptability in technically driven themes like social media/sentiment analysis. Although scholars across all stages exhibit synchronous trends, variations in the intensity of engagement reflect the influence of academic seniority on research depth and resource integration. These patterns offer a multi-dimensional perspective for understanding the development evolution of the discipline.

### 4.4 Evolution of research method usage in LIS scholars' academic careers

In this section, we address *RQ2* by exploring the evolution of research method usage in the academic careers of LIS scholars, both at the aggregate and individual levels. Building on the earlier analysis of the types of research methods used by scholars during their careers, this study examines the overall differences in method usage from 1990 to 2023. However, specific trends in the evolution may be obscured by factors such as the popularity of certain methods. Therefore, this subsection focuses on scholars who published their first paper between 1970 and 1979. This cohort was selected to minimize the generational effects of academic age differences on method usage and because scholars in this decade produced a higher volume of publications compared to other ten-year intervals, making them particularly valuable for analysis.

**Aggregate evolution of research method usage in LIS scholars' academic careers.** Figure 17 illustrates the evolving trends in research method usage among scholars who published their first paper between 1970 and 1979, as their academic age increased. Methods such as bibliometrics, content analysis, experiment, questionnaire and theoretical approach were widely used across different academic ages. Notably, experiment was more frequently employed when scholars were between 13 and 28 academic years old, while questionnaire became more prevalent after scholars reached 29 academic years of age. Over the course of their academic careers, scholars exhibited a trend toward greater diversity in the types of research methods they used as they aged.

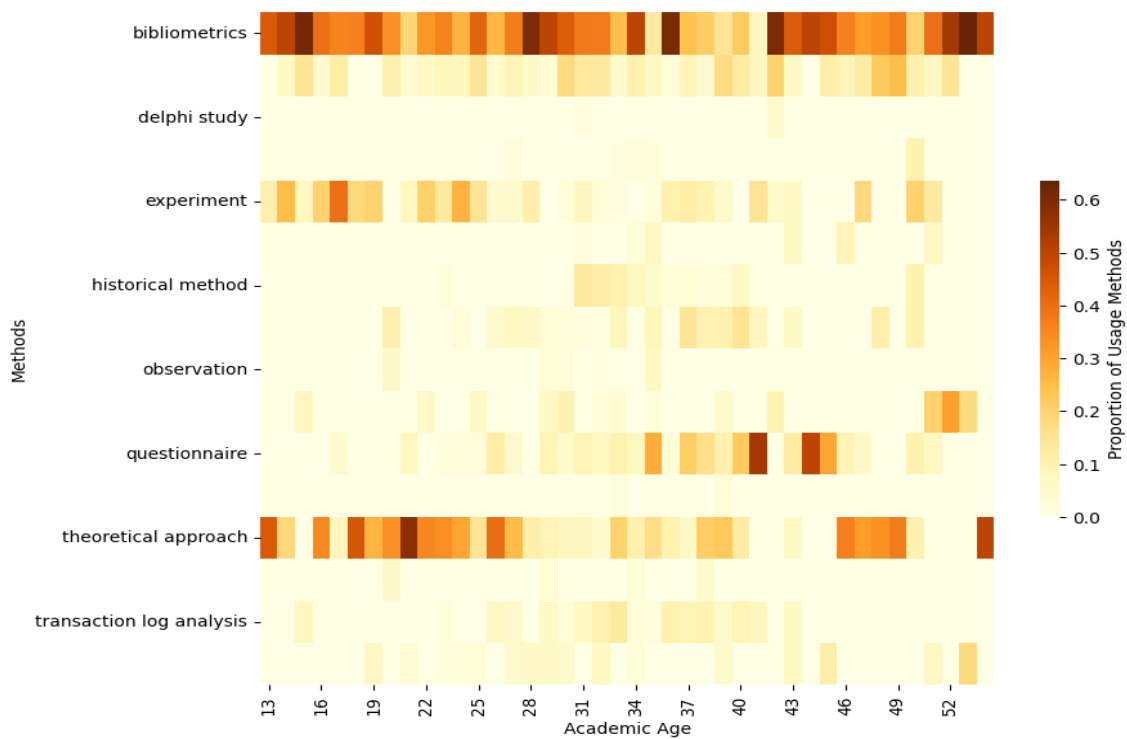

**Figure 17.** Evolution of research method usage among scholars at different career stages

To better demonstrate this relationship, we created an interactive heatmap.[3] This interactive graph collected data from the group of scholars whose earliest publication time was from 1970 to 1979. It can dynamically display the changes in the research methods used by scholars each year as their academic age increases. At the bottom of the interactive graph, there is a "Pause" button, which allows users to pause at any time to view the evolution of research methods in the academic careers of scholars in the LIS field in any specific year from 1990 to 2023. The detailed information of the selected scholars for this graph can be found in Table A of Appendix.

---
[3] https://jiayihao-njust.github.io/evolution/Method%20Evolution.html

**Individual evolution of research method usage in scholars' academic careers.** To further explore the characteristics of research method usage in scholars' academic careers, this study randomly selects four senior scholars and conducts a detailed analysis of their methodological evolution. Due to limited data availability, the analysis of these scholars' careers is based on their publications in the 15 selected journals between 1990 and 2023.For each scholar, the analysis focuses on the following aspects: the most frequently used research methods, the combination of methods employed, and the trends in changes to their research method usage over time.

Mike Thelwall is a male scholar whose first publication appeared in 2000. Over the course of his academic career, he has employed eight research methods, with the most frequently used being bibliometrics, webometrics, and content analysis. He has also utilized combined methods in his research, primarily pairing commonly used methods. During his early-career stage, he predominantly relied on webometrics. As he transitioned into the mid-career stage, his method repertoire expanded to include webometrics and content analysis, and he began incorporating combined methods into his research. In his senior-career stage, his most frequently used methods were bibliometrics, webometrics, and content analysis, with an increased reliance on combined methods. This demonstrates that his selection and use of research methods evolved in stages as he advanced in age and experience.

Amanda Spink is a female scholar whose first publication appeared in 1992. Throughout her academic career, she has employed ten research methods, with the most frequently used being transaction log analysis, questionnaire, content analysis, experiment, and theoretical approach. She has also employed combined methods in her research, including combinations of commonly used methods as well as pairings of common and less common methods. In some publications, she used up to four combined methods. Notably, the diversity of methods she employed remained consistent across different stages of her academic career, indicating her proficiency and habitual use of various methodologies to support her research endeavors.

Noa Aharony is a female scholar whose first publication appeared in 2006. Over her academic career, she has employed four research methods, with the most frequently used being questionnaire and content analysis. She has also utilized combined methods in her research, pairing commonly used methods with less common ones. During her early-career stage, her primary methods were questionnaire and content analysis. As she transitioned into the mid-career stage, the use of questionnaire increased significantly, while the use of content analysis declined relatively. She began employing combined methods and other methodologies during this period. In her senior-career stage, her most frequently used method was questionnaire. This suggests that, while her method choices exhibited a brief period of diversification during her mid-career stage, they ultimately stabilized. This stability may be attributed to the constraints of her research topics or her habitual preferences in method selection.

José Ortega is a male scholar whose first publication appeared in 2003. Throughout his academic career, he has employed four research methods, with the most frequently used being webometrics, content analysis, and bibliometrics. He has also utilized combined methods in his research, primarily pairing commonly used methods. During his early-career stage, his sole research method was webometrics. In his mid-career stage, his methodological choices evolved from webometrics to content analysis, then to a combination of content analysis and bibliometrics, and finally back to bibliometrics. In his senior-career stage, his most frequently used methods were content analysis and bibliometrics. This indicates a notable trend of method diversification during his mid-career stage.

From the evolution of research method usage among the four scholars described above, it is evident that during the mid-career stage, scholars exhibit a tendency to employ a diverse range of research methods, accompanied by an increase in publication output. The most frequently used research methods shift as scholars advance in age and experience, likely influenced by the popularity of certain methods and research topics during different periods. Throughout their academic careers, scholars experiment with various combinations of research methods, whether pairing commonly used methods or combining less common methods with popular ones. This reflects their flexibility and adaptability in applying research methods to their work.

## 5    Discussion

### 5.1    Research implications

**Theoretical implications.**  This study explores the evolution of research method usage in LIS scholars' careers and makes two unique contributions to the understanding of research method usage.

First, we combine the automatic categorization of research methods with scholars' academic careers to explore the relationship between scholars' academic age and research use. According to Bandura's (1986) social cognitive theory, the selection and regulation of human behavior fundamentally result from the dynamic interaction among individual cognition, behavior, and the environment. Scholars at different academic career stages may exhibit varying levels of personal cognitive understanding, which in turn influences their selection and application of research methods. In the early stages of their academic careers, scholars absorb method experiences from mentors and peers through observational learning. During this period, self-efficacy is gradually established through continuous experimentation, leading to a tendency to broadly combine and employ multiple research methods. As scholars progress in academic age and accumulate research experience, repeated practice fosters the formation of stable methodological habits. Self-regulatory mechanisms then encourage a focus on core research areas, resulting in an evolutionary pattern where method choices initially increase and subsequently decline. This process also highlights the long-term shaping effect of individual cognitive traits on preferences in research method selection.

Second, this study provides a comprehensive and dynamic overview of research method usage among LIS scholars from 1990 to 2023. It highlights innovative directions and the application of cutting-edge research methods within the LIS field, offering theoretical insights and guidance for disciplinary development and innovation.

**Practical Implications:** Focusing on the individual scholar level, this research examines the distinct patterns and characteristics in research method usage among scholars at different academic career stages. It aims to provide tailored guidance for researchers within the LIS field at various phases of their careers, moving beyond superficial insights gained merely from tracking senior scholars' publications.

For instance, while young scholars often look to the outputs of senior researchers—such as their published papers—to gain a preliminary understanding of commonly used methods, the findings of this study offer a more systematic framework for method learning. The research not only reveals which methods senior scholars employ but also clarifies how their method choices evolve over time. Furthermore, scholars may select different research methods based on their specific research interests and the broader historical context. Thus, young scholars need not passively imitate the current methods of senior researchers. Instead, they can proactively plan their own method development by aligning it with their research interests and career stage, engaging in targeted learning. This focused approach can more efficiently enrich a scholar's method repertoire, reduce trial and error, and accelerate professional growth.

From an institutional perspective, this study can inform the development of academic guidance programs that promote method diversity. Given the different method preferences observed across age groups, institutions can implement targeted initiatives. First, they can foster active collaboration among scholars of different generations, encouraging the application of diverse methods to research problems. Senior scholars can contribute their expertise in theoretical framing, while young scholars might introduce emerging techniques, such as machine learning for text classification, facilitating mutual learning. Second, institutions can help design structured guidance for diverse method choices. Given the variety of research methods in LIS, institutions can leverage data on method evolution to guide scholars in structured learning. For example, they might establish core method milestones tailored to different career stages.

## 5.2 Research limitations

Research on the usage evolution of research methods across academic careers among scholars in LIS still faces several key limitations. First, the study scope is restricted and potential sample bias exists. This research only utilizes publication data from 14 LIS journals spanning 1990 to 2023, failing to cover all core journals, conference proceedings, monographs and other academic output types in the field. Additionally, strict scholar screening criteria were applied, limiting the sample to senior scholars with an academic age of over 14 years and continuous publications every five years. This leads to inherent sample bias, meaning the observed research method usage patterns only reflect the characteristics of this specific senior scholar group and the selected journal scope, rather than the entire LIS scholarly community. Second, the adopted research method classification framework has inherent limitations. To ensure standardization and academic recognition, this study employs the classification system proposed by Chu and Ke (2017) for large-scale method categorization. However, this framework has non-negligible flaws: emerging technology-driven methods and interdisciplinary mixed methods cannot be fully adapted to it, often causing ambiguous coding or categorization errors. Third, the deep-seated influencing factors of research method selection are not thoroughly explored. This study only extracted and organized data on research method usage across scholars' academic careers, without probing into the underlying motivations for method selection. In-depth analysis of structural factors such as funding changes, journal review preferences and publishing policies was not conducted; meanwhile, individual factors including scholar collaboration, task division, academic background and personal research preferences also shape method choice, yet these were excluded from the analytical framework. Accordingly, these factors will be prioritized for in-depth investigation in follow-up research.

# 6      Conclusions and future research

We draw on data from 14 authoritative journals in the LIS field published between 1990 and 2023, selecting a subset of scholars to explore the evolution of research method usage in their academic careers.

Based on the above discussions, several conclusions can be drawn regarding the two research questions proposed in this study: Our findings indicate that the research methods commonly employed by scholars in the LIS field evolve with age and seniority, a shift likely influenced by factors such as prevailing methods and research trends across different periods. Throughout their academic careers, the diversity of research methods used by scholars initially increases and later declines. Scholars frequently experiment with combinations of methods—whether pairing commonly used techniques or integrating less conventional approaches with established ones—demonstrating flexibility in their methodological applications. Furthermore, scholars exhibit a clear developmental evolution: early-career researchers explore diverse methods, mid-career scholars reuse and adapt methods to broaden their thematic scope, and senior scholars achieve a balanced synergy among methodological expertise, thematic breadth, and research output. The adoption of research methods is shaped by multiple factors, including disciplinary evolution, technological advancements, and shifts in research priorities. Consequently, scholars selectively employ appropriate methodologies in response to contextual research demands across different periods.

In future work, we will focus on the following three priorities: First, expand data and sample coverage. This study will incorporate more diverse types of scholarly outputs in the LIS field, including additional core journals, conference proceedings, monographs, and preprints. Meanwhile, the scholar screening criteria will be relaxed to enhance the generalizability of the research findings. Second, optimize the research method classification system. In light of the developmental trends of emerging research methods and interdisciplinary approaches in LIS, this study will refine the method categorization criteria and add new classification entries for cutting-edge methods. Furthermore, localized and discipline-specific revisions and supplements will be made to the existing classification framework to better fit the field's unique characteristics. Finally, further explore the influencing factors of research method selection.This study will incorporate data on scholars' gender, research background, and collaboration networks into further analysis. It will thoroughly explore the factors shaping scholars' research method choices across their academic careers, and seek to establish causal relationships between these factors and method selection behaviors.

## Acknowledgments

This paper was supported by the National Natural Science Foundation of China (Grant No.72074113), Major Projects of National Social Science Fund (Grant No. 25&ZD298). This paper is an extended version of the ISSI 2025 paper "Hao, J., & Zhang, C. (2025). Trajectory of Research Method Usage in the Academic Careers of Scholars in the Library and Information Science. In: Proceedings of the 20th International Conference on Scientometrics and Informetrics (ISSI 2025), Yerevan, Armenia, 2025."

## References

Abramo, G., D'Angelo, C. A., & Murgia, G. (2016). The combined effects of age and seniority on research performance of full professors. Science and Public Policy, 43(3), 301–319.
Angelov, D. (2020). Top2Vec: Distributed Representations of Topics (No. arXiv:2008.09470). arXiv.
Ao, W., Lyu, D., Ruan, X., Li, J., & Cheng, Y. (2023). Scientific creativity patterns in scholars' academic careers: Evidence from PubMed. Journal of Informetrics, 17(4), 101463.
Aref, S., Zagheni, E., & West, J. (2019, November). The demography of the peripatetic researcher: Evidence on highly mobile scholars from the Web of Science. In International conference on social informatics (pp. 50-65). Cham: Springer International Publishing.
Azoulay, P., Fons-Rosen, C., & Zivin, J. S. G. (2019). Does Science Advance One Funeral at a Time? American Economic Review, 109(8), 2889–2920.
Badar, K., M. Hite, J., & F. Badir, Y. (2014). The moderating roles of academic age and institutional sector on the relationship between co-authorship network centrality and academic research performance. Aslib Journal of Information Management, 66(1), 38–53.
Bandura, A. . (1986). Social foundations of thought and action: a social cognitive theory. Journal of Applied Psychology, 12(1), 169.
Beltagy, I., Lo, K., & Cohan, A. (2019). SciBERT: A pretrained language model for scientific text. In K. Inui, J. Jiang, V. Ng, & X. Wan (Eds), Proceedings of the 2019 Conference on Empirical Methods in Natural Language Processing and the 9th

International Joint Conference on Natural Language Processing (EMNLP-IJCNLP) (pp. 3615–3620). Association for Computational Linguistics.

Bu, Y., Murray, D. S., Xu, J., Ding, Y., Ai, P., Shen, J., & Yang, F. (2018). Analyzing scientific collaboration with "giants" based on the milestones of career. Proceedings of the Association for Information Science and Technology, 55(1), 29–38.

Chan, H. F., & Torgler, B. (2020). Gender differences in performance of top cited scientists by field and country. Scientometrics, 125(3), 2421–2447.

Chowdhary, S., Gallo, L., Musciotto, F., & Battiston, F. (2024). Team careers in science: formation, composition and success of persistent collaborations. arXiv preprint arXiv:2407.09326.

Chu, H. (2015). Research methods in library and information science: A content analysis. Library & Information Science Research, 37(1), 36–41.

Coomes, O. T., Moore, T., Paterson, J., Breau, S., Ross, N. A., & Roulet, N. (2013). Academic Performance Indicators for Departments of Geography in the United States and Canada. The Professional Geographer, 65(3), 433–450.

Costas, R., Nane, G. F., & Lariviere, V. (2015). Is the Year of First Publication a Good Proxy of Scholars Academic Age?. In International Conference on Scientometrics & Informetrics (pp. 988-998). Retrieved from https://www.issi-society.org/proceedings/issi_2015/0988.pdf

Cui, H., Wu, L., & Evans, J. A. (2022). Aging scientists and slowed advance. arXiv preprint arXiv:2202.04044.

Ding, M., Zhou, C., Yang, H., & Tang, J. (2020). Cogltx: Applying bert to long texts. Advances in Neural Information Processing Systems, 33, 12792-12804.

Frandsen, T. F., & Nicolaisen, J. (2010). What is in a name? Credit assignment practices in different disciplines. Journal of Informetrics, 4(4), 608–617.

Győrffy, B., Csuka, G., Herman, P., & Török, Á. (2020). Is there a golden age in publication activity?—An analysis of age-related scholarly performance across all scientific disciplines. Scientometrics, 124(2), 1081–1097.

Hao, J., & Zhang, C. (2025). Trajectory of Research Method Usage in the Academic Careers of Scholars in the Library and Information Science. In: Proceedings of the 20th International Conference on Scientometrics and Informetrics (ISSI 2025), Yerevan, Armenia, 2025.

Hayman, R. & Smith, E. (2020). Mixed Methods Research in Library and Information Science: A Methodological Review. Evidence Based Library and Information Practice, 15(1), 106–125.

Heting Chu & Qing Ke. (2017). Research methods: What's in the name? Library & Information Science Research, 39(4), 284–294.

Järvelin, K., & Vakkari, P. (1990). Content Analysis of Research Articles in Library and Information Science. Library & Information Science Research, 12, 395-421.

Järvelin, K., & Vakkari, P. (1993). The evolution of library and information science 1965–1985: A content analysis of journal articles. Information Processing & Management, 29(1), 129–144.

Järvelin, K., & Vakkari, P. (2021). LIS research across 50 years: Content analysis of journal articles. Journal of Documentation, 78(7), 65–88.

Jia, T., Wang, D., & Szymanski, B. K. (2017). Quantifying patterns of research interest evolution. Nature Human Behaviour, 1(4), 0078.

Krauss, A. (2024). Science of science: A multidisciplinary field studying science. Heliyon, 10(17).

Kumar, S., & Ratnavelu, K. (2016). Perceptions of Scholars in the Field of Economics on Co-Authorship Associations: Evidence from an International Survey. PLOS ONE, 11(6), e0157633.

Liang, G., Hou, H., Ding, Y., & Hu, Z. (2020). Knowledge recency to the birth of Nobel Prize-winning articles: Gender, career stage, and country. Journal of Informetrics, 14(3), 101053.

Liao, C. H. (2017). Reopening the Black Box of Career Age and Research Performance. In J. Zhou & G. Salvendy (Eds.), Human Aspects of IT for the Aged Population. Applications, Services and Contexts (Vol. 10298, pp. 516–525). Springer International Publishing.

Liu, L., Jones, B. F., Uzzi, B., & Wang, D. (2023). Data, measurement and empirical methods in the science of science. Nature human behaviour, 7(7), 1046-1058.

Lou, W., Su, Z., He, J., & Li, K. (2021). A temporally dynamic examination of research method usage in the Chinese library and information science community. Information Processing & Management, 58(5), 102686.

Lund, B. D., & Wang, T. (2021). An analysis of research methods utilized in five top, practitioner-oriented LIS journals from 1980 to 2019. Journal of Documentation, 77(5), 1196–1208.

Milojević, S. (2012). How Are Academic Age, Productivity and Collaboration Related to Citing Behavior of Researchers? PLoS ONE, 7(11), e49176.

Nane, G. F., Larivière, V., & Costas, R. (2017). Predicting the age of researchers using bibliometric data. Journal of Informetrics, 11(3), 713–729.

Packalen, M., & Bhattacharya, J. (2019). Age and the Trying Out of New Ideas. Journal of Human Capital, 13(2), 341–373.

Palvia, P., Pinjani, P., & Sibley, E. H. (2007). A profile of information systems research published in Information & Management. Information & Management, 44(1), 1–11.

Perianes-Rodriguez, A., & Ruiz-Castillo, J. (2015). Within- and between-department variability in individual productivity: The case of economics. Scientometrics, 102(2), 1497–1520.

Robinson-Garcia, N., Costas, R., Sugimoto, C. R., Larivière, V., & Nane, G. F. (2020). Task specialization across research careers. eLife, 9, e60586.

Simoes, N., & Crespo, N. (2020). A flexible approach for measuring author-level publishing performance. Scientometrics, 122(1), 331–355.
Sugimoto, C., Sugimoto, T., Tsou, A., Milojevic, S., & Larivière, V. (2016). Age stratification and cohort effects in scholarly communication: A study of social sciences: Scientometrics, 109.
van den Besselaar, P., & Sandström, U. (2016). Gender differences in research performance and its impact on careers: A longitudinal case study. Scientometrics, 106, 143–162.
Wang, W., Yu, S., Bekele, T. M., Kong, X., & Xia, F. (2017). Scientific collaboration patterns vary with scholars' academic ages. Scientometrics, 112(1), 329–343.
Zeng, A., Shen, Z., Zhou, J., Fan, Y., Di, Z., Wang, Y., Stanley, H. E., & Havlin, S. (2019). Increasing trend of scientists to switch between topics. Nature Communications, 10(1), 3439.
Zhang, C., & Tian, L. (2023). Non-synchronism in global usage of research methods in library and information science from 1990 to 2019. Scientometrics, 128, 3981–4006.
Zhang, C., Tian, L., & Chu, H. (2023). Usage frequency and application variety of research methods in library and information science: Continuous investigation from 1991 to 2021. Information Processing & Management, 60(6), 103507.
Zhang, C., Zeng, J., & Zhao, Y. (2025). Is higher team gender diversity correlated with better scientific impact? Journal of Informetrics, 19(2), 101662.
Zhang, L., Qi, F., Sivertsen, G., Liang, L., & Campbell, D. (2024). Gender differences in the patterns and consequences of changing research directions in scientific careers. Quantitative Science Studies, 5(4), 882–905.
Zhang, Z., Tam, W., & Cox, A. (2021). Towards automated analysis of research methods in library and information science. Quantitative Science Studies, 2(2), 698–732.

# Appendix

**Table A.** Information on scholars with first publications between 1970 and 1979

| Earliest pub year | Author name | # publications (1990-2023) |
|---|---|---|
| 1970 | E. Michael Keen | 10 |
| 1970 | J. A. García | 49 |
| 1970 | Jaime A. Teixeira da Silva | 12 |
| 1970 | V.K. Singh | 21 |
| 1970 | W. W. Hood | 14 |
| 1971 | Anthony F. J. van Raan | 25 |
| 1971 | Barrie Gunter | 12 |
| 1971 | David Nicholas | 55 |
| 1971 | Michael E. D. Koenig | 12 |
| 1971 | Peter Vinkler | 26 |
| 1972 | Donald O. Case | 10 |
| 1972 | Peter Hernon | 14 |
| 1973 | Henry Small | 12 |
| 1973 | Ian Ruthven | 12 |
| 1973 | Jennifer Rowley | 56 |
| 1973 | M. H. Heine | 10 |
| 1973 | Peter Williams | 16 |
| 1974 | Mingyang Wang | 11 |
| 1975 | Gangan Prathap | 17 |
| 1976 | G.E. Gorman | 56 |
| 1976 | Maria Pinto | 76 |
| 1976 | R. Rada | 16 |
| 1977 | Birger Hjørland | 29 |
| 1977 | Howard D. White | 34 |
| 1977 | Hsin Hsin Chang | 11 |
| 1977 | Mark E. Rorvig | 20 |
| 1978 | Blaise Cronin | 36 |
| 1978 | Jin Zhang | 34 |
| 1978 | Leo Egghe | 226 |
| 1978 | Peter Willett | 27 |
| 1979 | Jin Ha Lee | 13 |
| 1979 | Nigel Ford | 23 |
| 1979 | Philip M. Davis | 16 |